\DocumentMetadata{}
\documentclass[sigconf]{acmart}
\usepackage{bm}
\usepackage{subfigure}
\usepackage{subcaption}
\usepackage{booktabs}
\usepackage{multirow}
\usepackage{multicol}
\usepackage{graphicx}
\usepackage{lipsum} 
\usepackage{algorithmic}
\usepackage[linesnumbered,ruled,noend]{algorithm2e} 
\usepackage{siunitx}
\usepackage{rotating}
\usepackage{enumitem}
\usepackage{wrapfig}

\usepackage{pifont}
\newcommand{\xmark}{\ding{55}}%
\theoremstyle{plain}

\theoremstyle{definition}

\theoremstyle{remark}

\DeclareMathOperator*{\argmin}{arg\,min}
\usepackage[textsize=small]{todonotes}
\usepackage{marginnote}
 \let\marginpar\marginnote

\AtBeginDocument{%
  \providecommand\BibTeX{{%
    \normalfont B\kern-0.5em{\scshape i\kern-0.25em b}\kern-0.8em\TeX}}}

\copyrightyear{2025}
\acmYear{2025}
\setcopyright{rightsretained}
\acmConference[WSDM '25]{Proceedings of the Eighteenth ACM International Conference on Web Search and Data Mining}{March 10--14, 2025}{Hannover, Germany}
\acmBooktitle{Proceedings of the Eighteenth ACM International Conference on Web Search and Data Mining (WSDM '25), March 10--14, 2025, Hannover, Germany}\acmDOI{10.1145/3701551.3703522}
\acmISBN{979-8-4007-1329-3/25/03}

\begin{document}

\title[SCONE]{SCONE: A Novel Stochastic Sampling to Generate Contrastive Views and Hard Negative Samples for Recommendation}
\author{Chaejeong Lee}
\authornote{Equal contribution.}
\authornote{This work was done partially while the author was with Yonsei University.}
\email{chaejeong.lee@kt.com}
\affiliation{%
  \institution{Korea Telecom}
  \city{Seoul}
  \country{South Korea}
}

\author{Jeongwhan Choi}
\email{jeongwhan.choi@yonsei.ac.kr}
\authornotemark[1]
\affiliation{%
  \institution{Yonsei University}
  \city{Seoul}
  \country{South Korea}
}

\author{Hyowon Wi}
\email{hyowon.wi@kaist.ac.kr}
\affiliation{%
  \institution{KAIST}
  \city{Daejeon}
  \country{South Korea}
}

\author{Sung-Bae Cho}
\email{sbcho@yonsei.ac.kr}
\affiliation{%
  \institution{Yonsei University}
  \city{Seoul}
  \country{South Korea}
}

\author{Noseong Park}
\authornote{Noseong Park is the corresponding author.}
\email{noseong@kaist.ac.kr}
\affiliation{%
  \institution{KAIST}
  \city{Daejeon}
  \country{South Korea}
}

\renewcommand{\shortauthors}{Lee, et al.}

\begin{abstract}
Graph-based collaborative filtering (CF) has emerged as a promising approach in recommender systems. Despite its achievements, graph-based CF models face challenges due to data sparsity and negative sampling. In this paper, we propose a novel \textbf{S}tochastic sampling for i) \textbf{CO}ntrastive views and ii) hard \textbf{NE}gative samples (SCONE) to overcome these issues. SCONE generates dynamic augmented views and diverse hard negative samples via a unified stochastic sampling approach based on score-based generative models. Our extensive experiments on 6 benchmark datasets show that SCONE consistently outperforms state-of-the-art baselines. SCONE shows efficacy in addressing user sparsity and item popularity issues, while enhancing performance for both cold-start users and long-tail items. Furthermore, our approach improves the diversity of the recommendation and the uniformity of the representations. The code is available at \url{https://github.com/jeongwhanchoi/SCONE}.
\end{abstract}



\maketitle

\begin{figure}[t!]
\centering
\subfigure[A stochastic process for contrastive views generation]{
\includegraphics[width=.98\columnwidth]{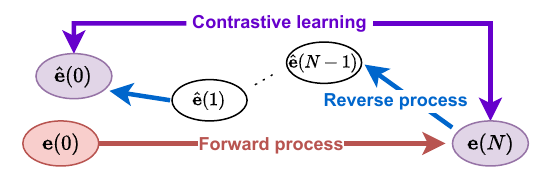}
\label{fig:intro_cl}
}
\subfigure[A stochastic positive injection for hard negative samples generation]{
 
\includegraphics[trim={0cm 0.1cm 0cm 0cm},clip, width=.98\columnwidth]{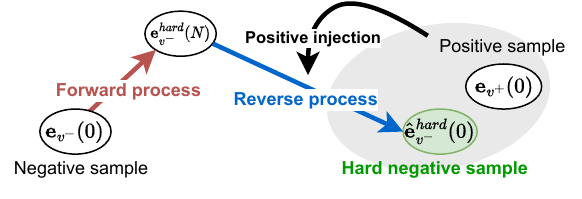}
\label{fig:intro_ns}
}
\caption{Our proposed user-item specific stochastic sampling method based on score-based generative models for contrastive learning and negative sampling. We generate contrastive views for CL with a stochastic process and hard negative samples with a stochastic positive injection.}
\label{fig:intro}
\end{figure}

\section{Introduction}
Recommender systems play a pivotal role in helping users discover relevant content or items from many choices in various domains, from product recommendation~\cite{linden2003amazon,hong2024svdae,shin2024attentive,lee2024gspcdr} and video recommendation~\cite{Covington2016YoutubeNet,ni2023content} to news recommendation~\cite{wu2019npa,lim2022airs}. Among various techniques, collaborative filterings (CFs) become popular solutions to predict user preferences, based on the rationale that users with similar interaction behaviors may share similar interests for items.

In recent years, the emergence of graph-based CF, which adopts graph convolution network (GCN), has gained much attention as a promising approach to model users' preferences for recommendations~\cite{Rianne2017GCMC,Rex2018pinsage,Wang19NGCF,He20LightGCN}. Graph-based models capture higher-order connectivity in user-item graphs, resulting in state-of-the-art performance~\cite{Mao21UltraGCN,choi2021ltocf,kong2022hmlet,hong2022timekit,choi2023bspm}. However, despite their success, graph-based CF models with implicit feedback suffer from data sparsity and negative sampling problems. 
Two main approaches have shown significant improvements in addressing these issues:
\begin{enumerate}[label=\roman*), leftmargin=*]
\item \textbf{Contrastive learning (CL)~\cite{Wu2021SGL,yu2022SimGCL,yu2022xsimgcl,xia2022HCCF}.} Compared to the entire interaction space, observed user-item interactions are sparse~\cite{bayer2017generic,he2016ups}. Most users only interact with a few items, and most items get a few interactions. This makse it hard to find hidden interactions in sparse data, and it is not enough to learn reliable node representations. Graph-based CF models adopt contrastive learning to solve the sparsity problem. This helps the model learn the general representations from unlabeled data space.

\item \textbf{Hard negative sampling~\cite{zhang2013dns,huang2021mixgcf, ding2020simplify,rendle2014improving, huang2020embedding,wu2023dimension,lai2023dens}.} When graph-based CF models are trained only with observed interactions, it is difficult for the model to predict unobserved interactions. The widely used BPR loss~\cite{rendle2009BPR} consists of observed and unobserved user-item pairs and is intended to prefer positive to negative pairs. However, the uniform negative sampling method has been proven to be less effective because it does not reflect the popularity of the item~\cite{He20LightGCN,Wang19NGCF}. Therefore, when training with BPR loss, it is a key point to define which samples are considered negative. Hard negative samples, which are close to positive samples, help the models learn the boundary between positive and negative samples.
\end{enumerate}
\begin{table}[t]
\small
\caption{Methods comparison for contrastive views and hard negative samples generation.}
\label{table:intro}
\begin{tabular}{cccc}
    \toprule
    Task & Method & Model Type & Domain \\
    \midrule
    \multirow{3}{*}{Contrastive Views} 
    & \cite{VGCL2023} & VAE & Recommendation \\
    & \cite{wu2023graph} & GAN & Graph \\
    & \cite{kim2023neural} & GAN & Computer Vision \\
    \cmidrule(lr){1-4}
    \multirow{3}{*}{Hard Negative} 
    & \cite{Wang2017IRGAN} & GAN & Recommendation \\
    & \cite{wang2021instance} & GAN & Computer Vision \\
    & \cite{li2022self} & GAN & Computer Vision \\
    \cmidrule(lr){1-4}
    Cont. Views \& Hard Neg. & SCONE & SGM & Recommendation \\
    \bottomrule
\end{tabular}
\end{table}

We observe that contrastive learning and hard negative sampling share a common factor: they generate tailored samples for specific purposes and use them during training. In recent years, various fields have adopted these approaches, especially using deep generative models~\cite{VGCL2023,wu2023graph,Wang2017IRGAN,wang2021instance,li2022self,kim2023neural}. In Table~\ref{table:intro}, we summarize existing methods for synthesizing contrastive views and hard negative samples with deep generative models. Since the generative models estimate the original dataset distribution, they can be used without information distortion.

Both contrastive views and hard negative samples potentially benefit from the capabilities of generative models, which means that they are technically similar to each other. This raises a natural question: \textbf{\emph{``given their similar requirements, can we integrate these two approaches into a single unified framework?''}} 

We answer this with a novel \textbf{S}tochastic sampling for \textbf{CO}ntrastive views and hard \textbf{NE}gative samples (\textbf{SCONE}), a user-item-specific stochastic sampling method based on score-based generative models (SGMs).
The overall design of our proposed method is as follows: 
\begin{enumerate}[label=\roman*), leftmargin=*]
    \item Following existing graph-based CF models, we adopt LightGCN as the graph encoder. Let $\mathbf{e}(0)$ be the final embedding of a specific user or item from LightGCN (see Fig.~\ref{fig:method_1}) --- given $\mathbf{e}(t)$, $t=0$ (resp. $t > 0$) typically means original (resp. noisy) samples in the SGM literature. We train a SGM using them.
    \item \textbf{Contrastive views generation via a stochastic process:} As in Fig.~\ref{fig:intro_cl}, we generate contrastive views with the forward and reverse processes of the SGM. The first view is a perturbed embedding $\mathbf{e}(N)$ through the forward process, and the second view is a generated embedding $\hat{\mathbf{e}}(0)$ from $\mathbf{e}(N)$ through the reverse process, where $N$ is an intermediate time. 
    \item \textbf{Hard negative samples generation via a stochastic positive injection:} As shown in Fig.~\ref{fig:intro_ns}, we generate a hard negative sample $\hat{\mathbf{e}}_{v^-}^{hard}(0)$ by injecting a positive sample $\mathbf{e}_{v^+}(0)$ during the reverse process. The stochastic positive injection generates hard negative samples close to positive ones.
\end{enumerate}

Our main contributions are summarized as follows:
\begin{enumerate}[label=\roman*), leftmargin=*]
    \item To solve the problems of data sparsity and negative sampling in graph-based CF models, SCONE addresses these challenges by considering that they are sampling tasks for the first time to our knowledge (Sec.~\ref{sec:scone}).
    \item We design a user-item specific stochastic sampling framework based on SGMs that generates dynamic contrastive views via stochastic processes (Sec.~\ref{subsec:method_cl}) and diverse hard negative samples via stochastic positive injection (Sec.~\ref{subsec:method_ns}).
    \item Extensive experiments on 6 benchmark datasets demonstrate the superiority and robustness of SCONE, particularly in handling user sparsity and item popularity issues (Secs.~\ref{subsec:rq1},~\ref{subsec:rq2}).
\end{enumerate}

\section{Preliminaries}
\subsection{Graph-based Collaborative Filtering}

Let $\mathcal{U}$ be the set of users, $\mathcal{I}$ be the set of items, and $\mathcal{O}^{+} = \{y_{ui}|u\in\mathcal{U}, i\in\mathcal{I}\}$ be the observed interactions, where $y_{ui}$ means that user $u$ has interacted with item $i$. The user-item relationships are constructed using a bipartite graph $\mathcal{G}=(\mathcal{V}, \mathcal{E})$, where the node set $\mathcal{V}=\mathcal{U}\cup\mathcal{I}$ is all users and items, and the edge set $\mathcal{E} = \mathcal{O}^{+}$ is the observed interactions.

Graph-based CF models adopt GCNs to capture users' preference on the item based on the neighborhood aggregation~\cite{Wang19NGCF,He20LightGCN,choi2021ltocf,kong2022hmlet,fan2022GTN,hu2024mgdcf}. In particular, LightGCN~\cite{He20LightGCN} simplifies GCNs by removing feature transformation and nonlinear activation as follows:
\begin{align}\label{eq:lgc}
\mathbf{E}^{(l+1)}=\tilde{\mathbf{A}}\mathbf{E}^{(l)},
\end{align} where $\mathbf{E}^{(0)} \in \mathbb{R}^{N \times D}$ is the trainable initial embedding with the number of nodes $N=|\mathcal{U}|+|\mathcal{I}|$ and $D$ dimensions of embedding, and $\mathbf{E}^{(l)}$ denotes the embedding at $l$-th layer. The normalized adjacency matrix is defined as $\tilde{\mathbf{A}} := \mathbf{D}^{-\frac{1}{2}}\mathbf{A}\mathbf{D}^{-\frac{1}{2}}$, where $\mathbf{A} \in \mathbb{R}^{N \times N}$ is the adjacency matrix and $\mathbf{D} \in \mathbb{R}^{N \times N}$ is the diagonal degree matrix.

Graph-based CF models utilize final embeddings by the weighted sum of the embeddings learned at each layer as follows: 
\begin{align}\label{eq:final_emb}
    \mathbf{e}_u = \sum_{l=0}^{L}\alpha_l\mathbf{e}_u^{(l)}, \quad \mathbf{e}_i = \sum_{l=0}^{L}\alpha_l\mathbf{e}_i^{(l)},
\end{align} where $\alpha_l$ is the importance of the $l$-th layer embedding in constructing the final embedding and $L$ is the number of layers. The inner product of user and item final embeddings is used to model prediction.
A widely used objective function is the pairwise Bayesian Personalized Ranking (BPR) loss:
\begin{align}
     \mathcal{L}_{BPR} = \sum_{(u,v^+,v^-)\in\mathcal{O}} {\ln{\sigma(\mathbf{e}_u \cdot \mathbf{e}_{v^-} - \mathbf{e}_u \cdot \mathbf{e}_{v^+} )}} + \lambda_2\|\Theta\|_2^2, \nonumber
\end{align} where $\mathcal{O}=\{(u,v^+,v^-) | (u,v^+)\in\mathcal{O}^+, (u,v^-)\in\mathcal{O}^-\}$ is the training data, $\mathcal{O}^-$ is the unobserved interactions, $\sigma(\cdot)$ is the sigmoid function, and $\lambda_2$ is to control the strengths of $L_2$.

\begin{figure}[t]
\centering
\includegraphics[trim={0cm 0.8cm 0cm 0.2cm},clip, width=.98\columnwidth]{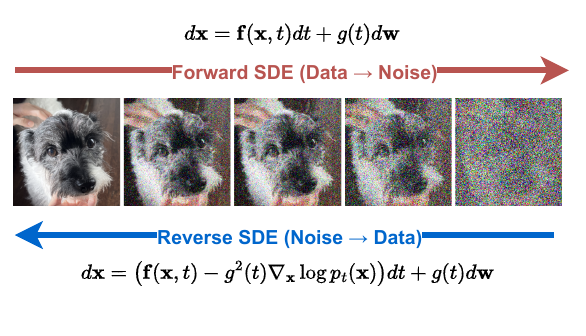}
\caption{The overall framework of score-based generative models. SGMs corrupt the input data with increasing noise and generate a new sample from the noisy data. 
}
\label{fig:preli_sgm}
\end{figure}

\subsection{Score-based Generative Models (SGMs)}

SGMs~\cite{song2021scorebased} effectively address the challenges of generative models. Fig.~\ref{fig:preli_sgm} introduces the overall framework of SGMs, which uses a stochastic differential equation (SDE) to model forward and reverse processes. The forward process involves adding noise to the initial input data, whereas the reverse process is to remove noise from the noisy data to generate a new sample.
SGMs define the forward SDE with the following It\^o SDE:
\begin{align}\label{eq:forward}
d\mathbf{x}=\mathbf{f}(\mathbf{x},t)dt + g(t)d\mathbf{w},
\end{align}where $\mathbf{f}(\mathbf{x},t) = f(t)\mathbf{x}$ and its reverse SDE is defined as follows:
\begin{align}\label{eq:reverse}
d\mathbf{x}=\big(\mathbf{f}(\mathbf{x},t)-g^2(t)\nabla_\mathbf{x} \log p_t(\mathbf{x})\big)dt + g(t)d\mathbf{w},
\end{align} which is a generative process. The function $f$ and $g$ are drift and diffusion coefficients, and $w$ is the standard Wiener process. According to the types of $f$ and $g$, SGMs are classified as i) variance exploding (VE), ii) variance preserving (VP), and ii) sub-variance preserving (Sub-VP) models. In our experiments, we only use VE SDE with the following $f$ and $g$:
\begin{align}
    f(t) = 0, \quad g(t) = \sqrt{\frac{d[\sigma^2(t)]}{dt}},
\end{align}where the noise scales $\sigma(t) = \sigma_{\text{min}}\Big(\frac{\sigma_{\text{max}}}{\sigma_{\text{min}}}\Big)^t$.

During the forward process, the score function $\nabla_\mathbf{x} \log p_t(\mathbf{x})$ is approximated by a score network $S_\phi(\mathbf{x}(t), t)$. Since the perturbation kernel $p(\mathbf{x}(t)|\mathbf{x}(0))$ can be easily collected, it allows us to calculate the gradient of the log transition probability $\nabla_{\mathbf{x}(t)} \log p(\mathbf{x}(t)|\mathbf{x}(0))$. Therefore, we can easily train a score network $S_\phi(\mathbf{x}(t), t)$ as follows:
\begin{align}
    \argmin_{\phi} \mathbb{E}_t \mathbb{E}_{\mathbf{x}(t)} \mathbb{E}_{\mathbf{x}(0)} \Big[\lambda(t) \big\|S_{\phi}(\mathbf{x}(t), t) -\nabla_{\mathbf{x}(t)} \log p(\mathbf{x}(t)|\mathbf{x}(0)) \big\|_2^2 \Big],\nonumber
\end{align} where $\lambda(t)$ is to control the trade-off between synthesis quality and likelihood. After training the score network, we can generate fake samples using only the reverse SDE.

\section{Existing approaches and limitations}

\subsection{Contrastive Learning for Recommendation}

Graph-based CFs face challanges such as data sparsity and cold-start problems. To address these, graph-based CFs use self-supervised learning (SSL), which augments various views and contrasts them to align node representations~\cite{Wu2021SGL,yu2022SimGCL,yu2022xsimgcl,choi2023qos,jing2023survey}. This can extract meaningful information from unlabeled interaction data.

SGL~\cite{Wu2021SGL}, the first method applying CL to graph-based CF, augments the user-item graph with structure perturbations. It uses InfoNCE loss~\cite{oord2018infonce} to contrast augmented views:
\begin{align}\label{eq:general_contrastive_loss}
\mathcal{L}_{CL} := \sum_{i \in \mathcal{B}} -\log\frac{\exp(\text{sim}(\mathbf{e}_i',\mathbf{e}_i'')/\tau)}{\sum_{j \in \mathcal{B}}\exp(\text{sim}(\mathbf{e}_i'\mathbf{e}_j'')/\tau)},
\end{align} where $i$, $j$ are a user and an item in a mini-batch $\mathcal{B}$ respectively, $\text{sim}(\cdot)$ is cosine similarity, $\tau$ is temperature, and $\mathbf{e}'$, $\mathbf{e}''$ are augmented node representations. 
The loss increases the alignment between the node representations of $\mathbf{e}_{i}'$ and $\mathbf{e}_{i}''$ nodes, viewing the representations of the same node $i$ as positive pairs. 
Simultaneously, it minimizes the alignment between the node representations of $\mathbf{e}_{i}'$ and $\mathbf{e}_{j}''$, viewing the representations of the different nodes $i$ and $j$ as negative pairs. 
However, it can result in a potential loss of detailed interaction information.

NCL~\cite{lin2022ncl} proposes a prototypical contrastive objective to capture the correlations between a user/item and its context. RDGCL~\cite{choi2023rdgcl} extracts contrastive views from reaction and diffusion processes to improve recommendation performance without the graph augmentation. SimGCL~\cite{yu2022SimGCL} pointed out that the graph augmentation method of SGL causes information loss and also plays only a trivial role in CL. It discards the graph augmentation and simply adds uniform noise to the embeddings, outperforming previous methods. However, SimGCL compromises the unique characteristics of each node on a graph by adding noises with the same scale.

\begin{figure}[t]
\centering
\subfigure[The overall model architecture consisting of LightGCN and SGM. We train our score network with the final embeddings of LightGCN.]{
\includegraphics[trim={0cm 0cm 0.5cm 0cm},clip, width=.98\columnwidth]{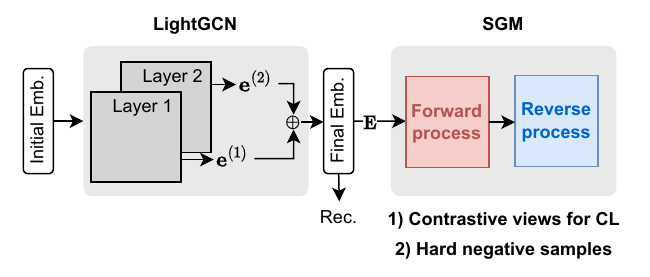}
\label{fig:method_1}
}
\subfigure[Contrastive views generation for a user/item embedding $\mathbf{e}(0)$ via a stochastic process. `F' and `Reverse' means the forward and reverse processes of SGMs, respectively. ]{
\includegraphics[trim={0cm 0cm 0.5cm 0cm},clip, width=.98\columnwidth]{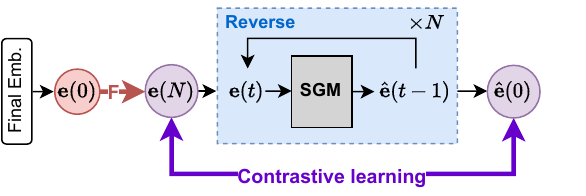}
\label{fig:method_2}
}
\subfigure[Hard negative sample generation via a stochastic positive injection given a user-item embedding triplet $(\mathbf{e}_u(0),\mathbf{e}_{v^+}(0), \mathbf{e}_{v^-}(0))$. The synthesized sample $\hat{\mathbf{e}}_{v^-}^{hard}(0)$ is a hard negative sample close to a positive sample.]{
\includegraphics[trim={0cm 0cm 1.3cm 0cm},clip, width=.98\columnwidth]{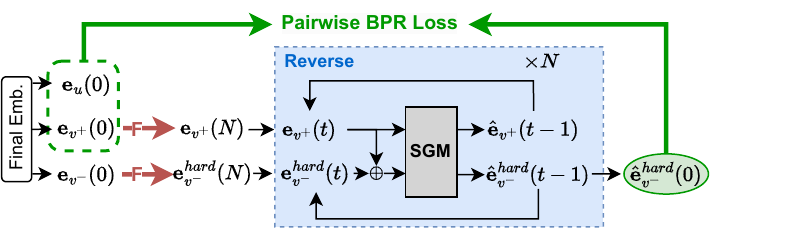}
\label{fig:method_3}
}
\caption{
The overall workflow of SCONE. We generate contrastive views and hard negative samples via stochastic sampling based on score-based generative models, and then utilize them for encoder training. 
}
\label{fig:method_all}
\end{figure}

\subsection{Negative Sampling Problem}
For graph-based CF models with implicit feedback, the negative sampling problem refers to the challenge encountered when training only with observed interactions. In most cases, user interactions such as purchasing products or clicking on items are considered positive signals, however, it can lead to difficulties when trying to predict unobserved interactions. 
Therefore, various negative sampling techniques have been explored to select proper negative signals from unobserved interactions. These approaches have shown significant success in improving the accuracy and diversity of recommendation systems~\cite{zhang2013dns, huang2021mixgcf, ding2020simplify, rendle2014improving, huang2020embedding, wu2023dimension,lai2023dens}.

A crucial aspect of negative sampling is defining hard negative samples to learn the boundary between positive and negative items. DNS~\cite{zhang2013dns} is a dynamic negative sampling strategy by selecting the negative item with the highest score. 
IRGAN~\cite{Wang2017IRGAN}, which is a GAN-based model, generates hard negative samples and combines with given positive samples.
MixGCF~\cite{huang2021mixgcf} is the state-of-the-art negative sampling method, which adopts positive mixing and hop mixing. In positive mixing, it utilizes the interpolation method by injecting positive information, and in hop mixing, it aggregates negative samples from selected hops of their neighbors. However, DNS and MixGCF are simplistic approaches for synthesizing hard negative samples only in a predetermined region, which can lead to limited diversity in them.

\section{Proposed Methods}\label{sec:scone}
We outline the 3 design goals of our proposed method:
\begin{enumerate}[label=\roman*), leftmargin=*]
      \item \textbf{Stochastic view generation.} We need to design a user-item specific stochastic sampling method to generate contrastive views for personalized recommendations.
      \item \textbf{Hard negative generation.} Our method should generate high-quality hard negative samples via stochastic positive injection to enhance BPR loss optimization and improve discriminative learning between user preferences.
      \item \textbf{Unified framework.} Our method should provide a single framework that can simultaneously handle contrastive view generation and hard negative sampling.
\end{enumerate} 
With these goals, we explore the overall architecture of our SCONE.

\subsection{Overall Model Architecture}\label{subsec:scone}
In Fig.~\ref{fig:method_1}, we show the overall architecture of SCONE. Our SCONE uses LightGCN as our graph encoder. Given the initial user and item embeddings $\Theta = \mathbf{E}^{(0)}$, LightGCN learns these embeddings. After that, we train a score-based generative model using the final embeddings $\mathbf{E}$, which is the weighted sum of the embeddings learned at all layers of LightGCN. The score function $\nabla_{\mathbf{e}} \log p_t(\mathbf{e})$ is approximated by a score network. The trained SGM is used in two ways: i) contrastive views generation for CL and ii) hard negative samples generation. More details will be described in Secs.~\ref{subsec:method_cl} and~\ref{subsec:method_ns}, respectively.

We design our score network modified from U-Net~\cite{ronneberger2015u}. We use 3 layers, 1 encoding block, and 1 decoding block linked by skip connection. The network architecture $S_{\phi}(\mathbf{e}(t), t)$ is as follows:
\begin{align}
&h_{0} = \mathtt{FC}_0(\mathbf{e}(t)), \\
&h_{1} = \mathtt{FC}_1^2(\mathtt{tanh}(\mathtt{FC}_1^1(h_{0}))\oplus\mathtt{FC}_1^t(t_\text{emb})), \\
&h_{2} = \mathtt{tanh}(\mathtt{FC}_2(h_{1})), \\
&h_{3} =\mathtt{FC}_3^2(\mathtt{tanh}(\mathtt{FC}_3^1(h_{2}\odot h_{1}))\oplus\mathtt{FC}_3^t(t_{\text{emb}})), \\
&S_{\phi}(\mathbf{e}(t), t) = \mathtt{FC}_4(h_{3}),
\end{align}
where $\mathtt{FC}$ is a fully connected layer, $\odot$ is a concatenation operator, $h_{i}$ is an $i^{th}$ hidden vector, and $\oplus$ is an element-wise addition. The encoding and decoding blocks are symmetric, where $\text{dim}(h_0) = \text{dim}(h_3)$ and $\text{dim}(h_1) = \text{dim}(h_2)$, and the output of the model has the same dimensionality $D$ as the input $\mathbf{e}$. The time embedding $t_{emb}$ is defined as follows:
\begin{equation}
    t_\text{emb} = \mathtt{FC}_\text{emb}^2(\mathtt{tanh}(\mathtt{FC}_\text{emb}^1(\mathtt{Emb}(t)))),
\end{equation}where $\mathtt{Emb}$ is a sinusoidal positional embedding~\cite{vaswani2017attention}.

\subsection{User-Item Specific Stochastic Sampling}\label{subsec:specific} 

We propose a user-item specific sampling scheme for personalized recommendations. As shown in Figs.~\ref{fig:method_2} and~\ref{fig:method_3}, we perturb the final embedding $\mathbf{e}(0)$ to an intermediate time $t=N$, then generate a sample $\hat{\mathbf{e}}(0)$ from $\mathbf{e}(N)$ using the reverse process of SGM. This retains specific characteristics of users and items, which are applicable to both contrastive views and hard negative sample generation. We set an intermediate time $N=10$ in our experiments, and the impact of $N$ is discussed in Sec.~\ref{subsec:rq4}. 

\subsection{Contrastive Views Generation via a 
Stochastic Process}\label{subsec:method_cl}

We define two contrastive views for CL with SGM (See Fig.~\ref{fig:method_2}):
\begin{align}
    \mathbf{e}' = \mathbf{e}(N),\quad
    \mathbf{e}'' = \hat{\mathbf{e}}(0),
\end{align}
where the first view $\mathbf{e}'$ is a perturbed embedding $\mathbf{e}(N)$ using the forward process with Eq.~\eqref{eq:forward}, and the second view $\mathbf{e}''$ is a generated embedding $\hat{\mathbf{e}}(0)$ from $\mathbf{e}(N)$ using the reverse process with Eq.~\eqref{eq:reverse}. The detailed stochastic sampling process for the generation of contrastive views is in Alg.~\ref{alg:cl}. We sample an intermediate noisy vector $\mathbf{e}(N)$ with the forward process and convert it into a synthesized sample $\hat{\mathbf{e}}(0)$ through $N$ steps. The stochastic process of the SGM does not converge to the original but to another sample.

\begin{algorithm}[t]
   \small
   \caption{Contrastive views generation}
   \label{alg:cl}
\begin{algorithmic}[1]
   \REQUIRE $N$: Number of sampling steps for the reverse process
   \STATE $\mathbf{e}(N) \sim \mathcal{N}\big(\mathbf{e}(0), [\sigma^2(N)-\sigma^2(0)]\mathbf{I}\big)$
   \FOR {$i=N-1$ to $0$}
   \STATE $\mathbf{e}(i) \gets \mathbf{e}(i+1) + \big(\sigma^2(i+1)-\sigma^2(i)\big)\cdot S_{\phi}\big(\mathbf{e}(i+1), \sigma(i+1)\big)$
   \STATE $z \sim \mathcal{N}(\mathbf{0}, \mathbf{I})$
    \STATE $\mathbf{e}(i) \gets \mathbf{e}(i)+\sqrt{\sigma^2(i+1)-\sigma^2(i)} \cdot z$
   \ENDFOR
   \RETURN{$\mathbf{e}(N)$ and $\hat{\mathbf{e}}(0)$}
\end{algorithmic}
\end{algorithm}

\begin{algorithm}[t]
   \small 
   \caption{Hard negative samples generation}
   \label{alg:ns}
\begin{algorithmic}[1]
   \REQUIRE $N$: Number of sampling steps for the reverse process 
  \STATE $\mathbf{e}_{v^-}^{hard}(N) \sim \mathcal{N}\big(\mathbf{e}_{v^-}(0), [\sigma^2(N)-\sigma^2(0)]\mathbf{I}\big)$
 \STATE $\mathbf{e}_{v^+}(N) \sim \mathcal{N}\big(\mathbf{e}_{v^+}(0), [\sigma^2(N)-\sigma^2(0)]\mathbf{I}\big)$
   \FOR {$i=N-1$ to $0$}

    \STATE $\mathbf{e}_{v^-}^{hard}(i+1) = w \cdot \mathbf{e}_{v^-}^{hard}(i+1) + (1-w) \cdot \mathbf{e}_{v^+}(i+1)$
    \STATE $ \mathbf{e}(i+1) \gets \mathrm{CONCAT} \big(\mathbf{e}_{v^-}^{hard}(i+1), \mathbf{e}_{v^+}(i+1)\big)$
   \STATE $\mathbf{e}(i) \gets \mathbf{e}(i+1) + \big(\sigma^2(i+1)-\sigma^2(i)\big) \cdot S_{\phi}\big(\mathbf{e}(i+1), \sigma(i+1)\big)$
   \STATE $z \sim \mathcal{N}(\mathbf{0}, \mathbf{I})$
    \STATE $\mathbf{e}(i) \gets \mathbf{e}(i)+\sqrt{\sigma^2(i+1)-\sigma^2(i)}\cdot z$
    \STATE Split $\mathbf{e}(i)$ into two vectors $\mathbf{e}_{v^-}^{hard}(i)$ and $\mathbf{e}_{v^+}(i)$.
   \ENDFOR
   \RETURN{$\hat{\mathbf{e}}_{v^-}^{hard}(0)$}
\end{algorithmic}
\end{algorithm}

\subsection{Hard Negative Samples Generation via a Stochastic Positive Injection}\label{subsec:method_ns}

To enhance the learning of the boundary between positive and negative samples, we propose a stochastic positive injection method to generate hard negative samples close to positive samples. Fig.~\ref{fig:method_3} shows the process of generating hard negative samples. Firstly, we use the random sampling method in the BPR loss, where positive and negative samples are sampled from a uniform distribution. After sampling a triplet input $(u,v^+,v^-)$, we sample intermediate noisy vectors $\mathbf{e}_{v^-}^{hard}(N)$ and $\mathbf{e}_{v^+}(N)$ with the forward process. Then, positive samples are continuously injected into negative samples during the stochastic sampling process of SGM. The stochastic positive injection is defined as follows:
\begin{align}
    \mathbf{e}_{v^-}^{hard}(t) = w\cdot \mathbf{e}_{v^-}^{hard}(t) + (1-w)\cdot\mathbf{e}_{v^+}(t)
\end{align}where $w$ is hyperparameters to regulate the weight of stochastic positive injection. Alg.~\ref{alg:ns} shows the detailed hard negative samples generation process. We define a hard negative sample $\mathbf{e}_{v^-}^{hard}(t)$ by injecting a positive sample $\mathbf{e}_{v^+}(t)$ for $N$ steps, and finally, we generate a hard negative sample $\hat{\mathbf{e}}_{v^-}^{hard}(0)$. These synthesized samples cover the
entire embedding region rather than a predefined point.

\subsection{Training Algorithm}
Alg.~\ref{alg:training} outlines the training process of SCONE.
We jointly train the parameters $\Theta$ and $\phi$ of the graph encoder and SGM end-to-end, respectively.
The objective functions for approximating $\Theta$ and $\phi$ are as follows:
\begin{align}\label{eq:encoder_loss}
    \mathcal{L}_{\Theta} &= \mathcal{L}_{BPR} + \lambda_1\mathcal{L}_{CL},
\end{align}
\begin{align}\label{eq:sgm_loss}
    \mathcal{L}_{\phi} &= \mathbb{E}_t \mathbb{E}_{\mathbf{e}(t)} \mathbb{E}_{\mathbf{e}(0)} \Big[\lambda(t) \|S_{\phi}(\mathbf{e}(t), t) -\nabla_{\mathbf{e}(t)} \log p(\mathbf{e}(t)|\mathbf{e}(0)) \|_2^2 \Big],
\end{align}
where $\lambda_1$ is hyperparameters to control the CL loss. We adopt CL loss as InfoNCE~\cite{oord2018infonce}. $\mathcal{L}_{BPR}$ and $\mathcal{L}_{CL}$ are as follows:
\begin{align}\label{eq:bpr_loss}
     \mathcal{L}_{BPR} = \sum_{(u,v^+,v^-)\in\mathcal{O}} {\ln{\sigma(\mathbf{e}_u \cdot \hat{\mathbf{e}}_{v^-}^{hard}(0) - \mathbf{e}_u \cdot \mathbf{e}_{v^+} )}} + \lambda_2\|\Theta\|_2^2,
\end{align}
\begin{align}\label{eq:cl_loss}
    \mathcal{L}_{CL} = \sum_{i \in \mathcal{B}} -\log\frac{\exp(\text{sim}(\mathbf{e}_i',\mathbf{e}_i'')/\tau)}{\sum_{j \in \mathcal{B}}\exp(\text{sim}(\mathbf{e}_i',\mathbf{e}_j'')/\tau)},
\end{align}where $\hat{\mathbf{e}}_{v^-}^{hard}(0)$ is a generated hard negative sample by SGM (cf. Line~\ref{a:ns}
of Alg.~\ref{alg:training}), and $\mathbf{e}'$ and $\mathbf{e}''$ are contrastive views with the stochastic process of SGM (cf. Line~\ref{a:cl} of Alg.~\ref{alg:training}).

\begin{algorithm}[t]
   \small 
   \caption{The training algorithm of SCONE}
   \label{alg:training}
\begin{algorithmic}[1]
   \REQUIRE Training set $\{(u,v^+,v^-)\}$, Graph encoder $f(\cdot)$, Score network $S(\cdot)$
   \STATE Initialize $\Theta$ and $\phi$
   \WHILE{not converged}
       \STATE Sample a batch of user-item pairs $(u,v^+,v^-)$.
        \STATE Get the user and item embeddings by encoder $f(\cdot)$.
        \STATE Generate augmented views $\mathbf{e}'$ and $\mathbf{e}''$ by Alg.~\ref{alg:cl}. \label{a:cl}
        \STATE Generate a hard negative sample $\hat{\mathbf{e}}_{v^-}^{hard}(0)$ by Alg.~\ref{alg:ns}. \label{a:ns}
        \STATE Calculate $\mathcal{L}_{\Theta}$ with $\mathcal{L}_{BPR}$ and $\mathcal{L}_{CL}$ by Eq.~\eqref{eq:encoder_loss}.
        \STATE Calculate $\mathcal{L}_{\phi}$ with Eq.~\eqref{eq:sgm_loss}.
       \STATE Update $\Theta$ and $\phi$.
   \ENDWHILE
   \RETURN{$\Theta$ and $\phi$}
\end{algorithmic}
\end{algorithm}

\begin{table}[t]
\caption{Comparison of existing methods that focuses on i) how to generate contrastive views for CL and ii) how to define negative samples}
\setlength\tabcolsep{3pt}
\label{table:preli}
\begin{center}
\begin{small}
\begin{tabular}{lcc}
\toprule
Model & Contrastive Views & Negative Samples \\
\midrule
LightGCN~\cite{He20LightGCN} & \xmark & Random sampling\\
SGL~\cite{Wu2021SGL} & Perturbing graph structure & Random sampling \\
SimGCL~\cite{yu2022SimGCL} & Injecting noises into embeddings & Random sampling \\
DNS~\cite{zhang2013dns} & \xmark & Promising candidate \\
MixGCF~\cite{huang2021mixgcf}  & \xmark & Linear interpolation\\
\midrule
SCONE & \multicolumn{2}{c}{Stochastic sampling with SGMs} \\
\bottomrule
\end{tabular}
\end{small}
\end{center}
\end{table}

\begin{table*}[t]
\caption{Performance comparison with state-of-the-art models. The best results are highlighted in bold face, and the second best results are with underline.}
\label{table:main}
\begin{center}
\begin{tabular}{lccccccccccccc}
\toprule
\multirow{2}{*}{Dataset} & \multirow{2}{*}{Metric} & GCN-based && \multicolumn{3}{c}{Contrastive Learning} && \multicolumn{2}{c}{Negative Sampling} && \multirow{2}{*}{\textbf{SCONE}} & \multirow{2}{*}{\textit{Imp.}}\\
\cmidrule{3-3} \cmidrule{5-7} \cmidrule{9-10}
&& LightGCN && SGL & NCL & SimGCL && DNS &  MixGCF && \\
\midrule
\multirow{2}{*}{Douban} & Recall@20 & 0.1474 && 0.1728 & 0.1638 & 0.1780  && 0.1645 & \underline{0.1784} && \textbf{0.1815} & \textit{1.70\%} \\
                             & NDCG@20   & 0.1240 && 0.1510 & 0.1403 & 0.1567  && 0.1393 &\underline{0.1576} && \textbf{0.1611} & \textit{2.79\%} \\
\midrule  	 	 	 	 	 	 	 	 	 	 	 
\multirow{2}{*}{Gowalla}     & Recall@20 & 0.2086 && 0.2223 & 0.2235 & \underline{0.2269}  && 0.2233 & 0.2147 && \textbf{0.2295} & \textit{1.15\%} \\
                             & NDCG@20   & 0.1264 && 0.1349 & 0.1353 & \underline{0.1386}  && 0.1358 & 0.1295 && \textbf{0.1409} & \textit{1.66\%} \\
\midrule
\multirow{2}{*}{Tmall}       & Recall@20 & 0.0686 && 0.0754 & 0.0700 & \underline{0.0898}  && 0.0860 & 0.0819 && \textbf{0.0908} & \textit{1.08\%}\\
                             & NDCG@20   & 0.0477 && 0.0532 & 0.0492 & \underline{0.0643}  && 0.0600 & 0.0570 && \textbf{0.0654} & \textit{1.73\%} \\
\midrule                             
\multirow{2}{*}{Yelp2018}    & Recall@20 & 0.0588 && 0.0669 & 0.0654 & \underline{0.0718} && 0.0671 &  0.0703 && \textbf{0.0721} & \textit{0.39\%} \\
                             & NDCG@20   & 0.0485 && 0.0552 & 0.0540 & \underline{0.0593}  && 0.0551 & 0.0576 && \textbf{0.0594} & \textit{0.27\%} \\
\midrule                             
\multirow{2}{*}{Amazon-CDs}  & Recall@20 & 0.1283 && 0.1550 & 0.1483 & \underline{0.1584}  && 0.1545 &  0.1542 && \textbf{0.1586} & \textit{0.13\%} \\
                             & NDCG@20   & 0.0779 && 0.0986 & 0.0940 & \underline{0.1006}  && 0.0972 &  0.0960 && \textbf{0.1009} & \textit{0.30\%} \\
\midrule                             
\multirow{2}{*}{ML-1M}       & Recall@20 & 0.2693 && 0.2657 & 0.2777 & 0.2868  && 0.2791 & \underline{0.2897} && \textbf{0.2910} & \textit{0.45\%} \\
                             & NDCG@20   & 0.3015 && 0.3027 & 0.3150 & 0.3230  && 0.3171 & \textbf{0.3287} && \underline{0.3244} & \textit{-1.31\%} \\
\bottomrule
\end{tabular}
\end{center}
\end{table*}

\subsection{Discussion: Relation to Other Methods}
Table~\ref{table:preli} compares existing approaches for generating contrastive views and negative samples. We analyze SCONE in relation to these methods in terms of pros and cons, and space and time complexity.

\subsubsection{Contrastive views}
Diverse users and items exhibit distinctive characteristics, and view generation techniques should be customized for each user.
SCONE generates contrastive views using a stochastic process with score-based generative models. It is capable of approximating data distributions without information distortion. In particular, it can sample with high diversity while preserving specific characteristics of each node for personalized recommendations due to the user-item specific stochastic sampling. However, SGL and SimGCL are exposed to information distortions. SGL perturbs the inherent characteristic of the graph through edge/node dropouts. In addition, SimGCL distort the individual characteristic of each node by adding noises.

\subsubsection{Negative sampling} 
SCONE generates hard negative samples covering the entire embedding region using stochastic positive injection. 
However, LightGCN, SGL, and SimGCL adopt a random negative sampling (RNS) method, which randomly selects unobserved interaction signals. This approach often fails to capture the popularity of the item. Furthermore, DNS and MixGCF confine the selection of negative samples into a limited space. For DNS, promising candidate items are predefined in a discrete space, and for MixGCF, the sampled hard negative sample is a linear interpolation between the positive and negative items.

\subsubsection{Space and time complexity}
Regarding the model parameters, LightGCN, SGL, SimGCL, DNS, and MixGCF only require initial embeddings $\Theta = \mathbf{E}$, while SCONE needs additional parameters $\phi$ for the score network. However, the number of the parameters $\Theta$ is 9 times larger than that of $\phi$ in Douban, and 24 times in Tmall. Therefore, it does not significantly increase the space complexity.

In terms of time complexity, SGL and SimGCL require 3 iterations of GCNs to generate the final embeddings for recommendation and 2 augmented embeddings for CL. DNS and MixGCF need only one iteration. However, they need considerable time to define hard negative samples. For DNS, it should sort the highest negative items after referring to the currently trained recommendation model during training, and for MixGCF, it uses a positive and hop mixing module by the linear combination and hop-grained scheme. Our SCONE also requires only one iteration of GCNs. However, SCONE incurs overheads for generating contrastive views and hard negative samples. However, this can be mitigated by adjusting the number of steps, $N$. Sec.~\ref{sec:runtime} provides a detailed runtime comparison. SCONE takes a longer time compared to graph-based CL methods, but is faster than negative sampling methods.

\section{Experiments}\label{sec:exp}
In this section, we introduce our experimental results. To show the superiority of SCONE and clarify its effectiveness, we answer the following research questions:
\begin{itemize}
    \item \textbf{RQ1:} How does SCONE perform as compared with various state-of-the-art models?
    \item \textbf{RQ2:} Is the robustness of SCONE, with respect to user sparsity and item popularity, superior to the baselines?
    \item \textbf{RQ3:} Do all proposed methods improve the effectiveness?
    \item \textbf{RQ4:} What is the influence of different settings on SCONE?
\end{itemize}

\subsection{Experimental Settings}
\subsubsection{Datasets and Evaluation Metrics}
We conduct experiments on 6 benchmark datasets: Douban, Gowalla, Tmall, Yelp2018, Amazon-CDs, and ML-1M~\cite{yu2021douban, He20LightGCN, cai2023lightgcl, harper2015movielens, yu2022SimGCL}. Detailed statistics of the datasets are provided in Appendix~\ref{app:exp-detail}. Following \citet{yu2022SimGCL}, we split the datasets into train, validation, and test set with a ratio of 7:1:2. We use Recall@20 and NDCG@20 to evaluate the performance of the top-K recommendation.

\subsubsection{Baselines}
We compare SCONE with 6 baselines: LightGCN~\cite{He20LightGCN}, SGL~\cite{Wu2021SGL}, NCL~\cite{lin2022ncl}, SimGCL~\cite{yu2022SimGCL}, DNS~\cite{zhang2013dns}, and MixGCF~\cite{huang2021mixgcf}.
To make a fair comparison, we use hyperparameter settings from the original papers. 
We use SELFRec\footnote{https://github.com/Coder-Yu/SELFRec} library~\cite{yu2023self} for implementation.
For SCONE, we provide detailed hyperparameter settings and the best configurations in Appendix~\ref{app:exp-detail}.

\subsection{Performance Comparison (RQ1)}\label{subsec:rq1}
As shown in Table~\ref{table:main}, we summarize the overall performance of 6 baselines on the 6 datasets in terms of Recall@20 and NDCG@20. The results show that SCONE achieves the best performance in almost all cases. On Douban dataset, SCONE improves Recall@20 by 1.70\% and NDCG@20 by 2.79\% compared to MixGCF. For sparser datasets such as Gowalla and Amazon-CDs, SCONE shows siginifant improvements, and this indicates its effectiveness in addressing data sparsity issues. In negative sampling methods, DNS and MixGCF are hard to determine superiority. SimGCL shows the best results among the graph-based CL models, but it does not outperform our SCONE. These results show the effectiveness of our novel approach of generating augmented views and hard negative samples through stochastic sampling.

\subsection{Robustness of SCONE (RQ2)}\label{subsec:rq2}

\subsubsection{User sparsity}
\begin{figure}[t!]
\centering
\subfigure[Gowalla]{
\includegraphics[width=.47\columnwidth]{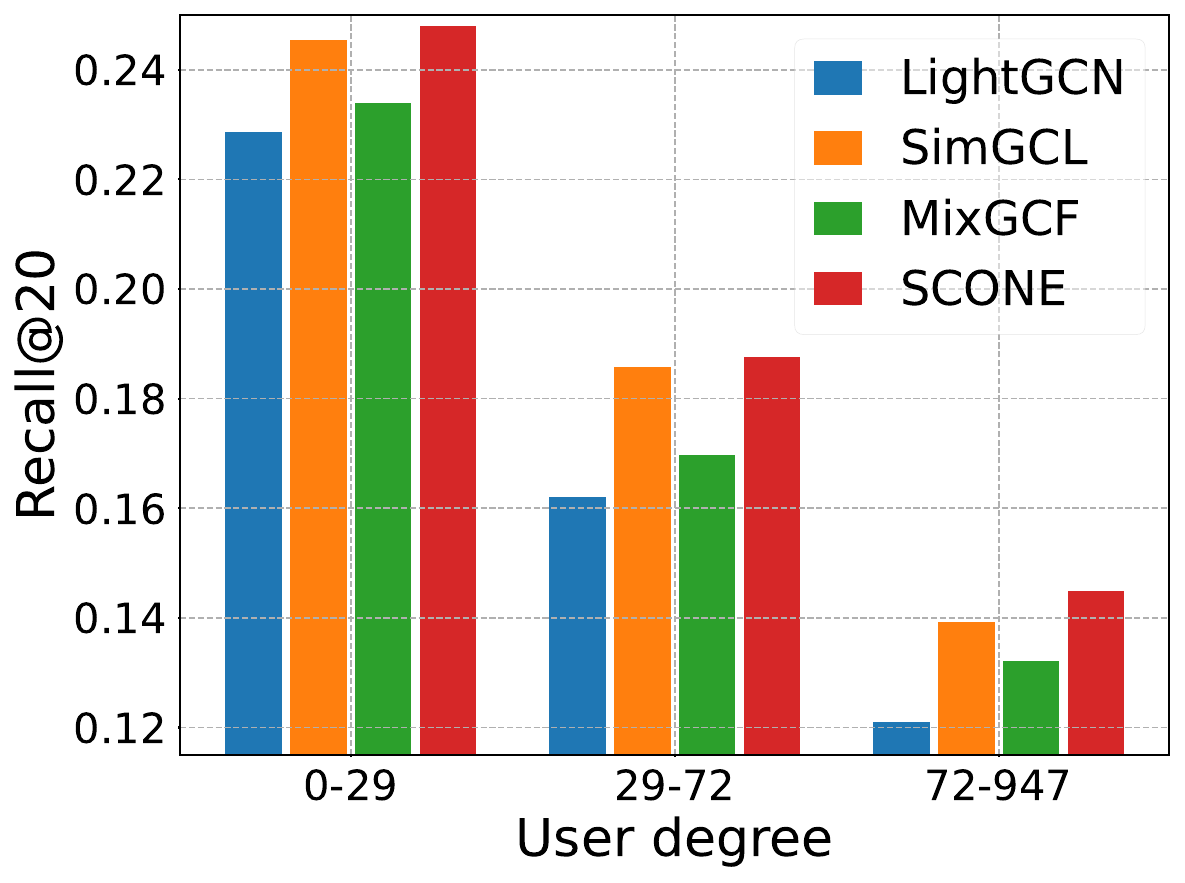}
\label{fig:exp_user_gowalla}
}
\subfigure[Tmall]{
\includegraphics[width=.47\columnwidth]{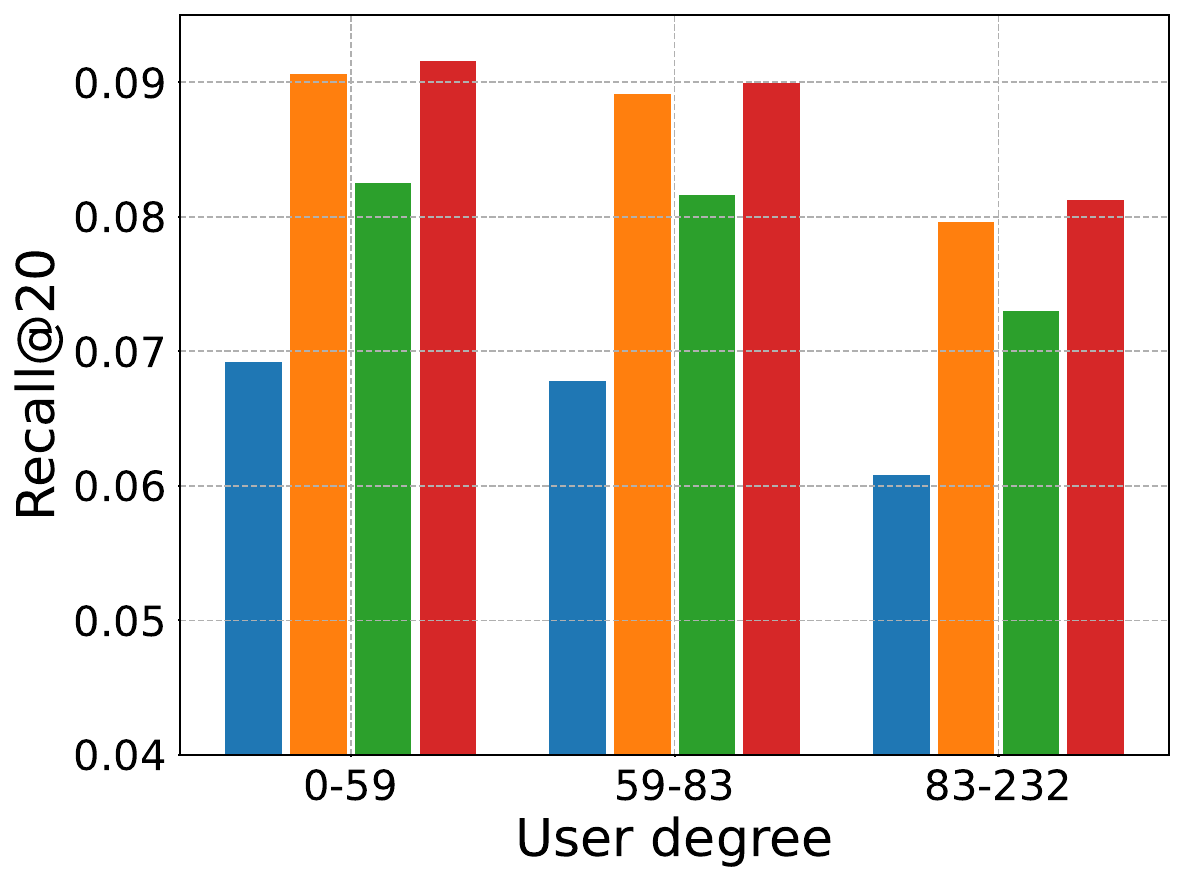}
\label{fig:exp_user_tmall}
}
\caption{Performance comparison over different user groups. More results are in oAppendix~\ref{app:robustness}.}\label{fig:exp_user}
\vspace{-0.5em}
\end{figure}

To evaluate the ability to alleviate user sparsity, we categorize users into three subsets by their interaction degree: the lowest 80\% users with the fewest number of interactions, the range from the bottom 80\% to 95\%, and the top 5\%. As shown in Fig.~\ref{fig:exp_user}, SCONE consistently outperforms the other baselines in almost all groups. Especially, SCONE shows good performance for extremely sparse user groups in all datasets.

\subsubsection{Item popularity}
Similar to user sparsity analysis, we categorize items into three groups based on item popularity. We calculate the Recall@20 value that each group contributes, which is called decomposed Recall@20. The total Recall@20 is equal to the sum of the values of the 3 groups. In Fig.~\ref{fig:exp_item}, our model shows performance improvement over other baselines in an item group with low popularity, demonstrating robustness to item popularity.

\subsubsection{Uniformity}
\begin{figure}[t]
\centering
\subfigure[Douban]{
\includegraphics[width=.47\columnwidth]{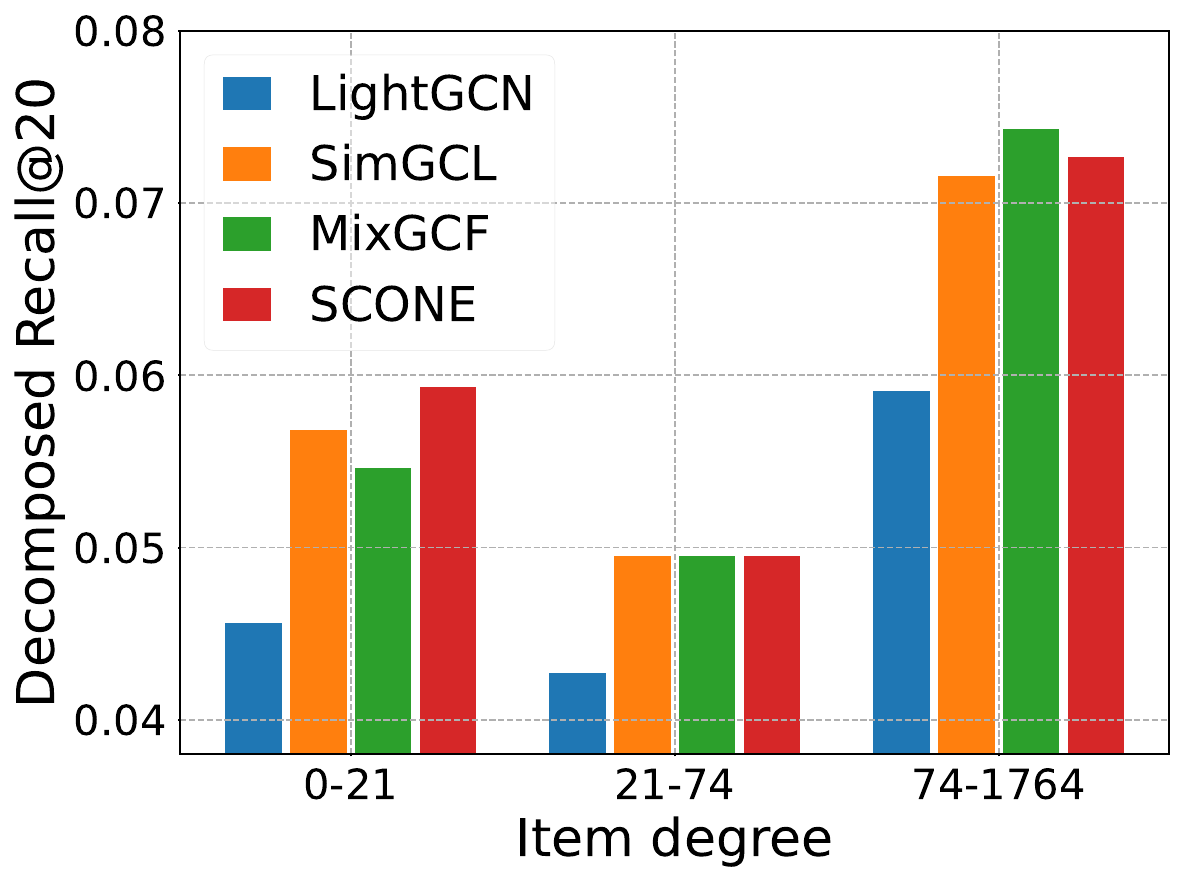}
\label{fig:exp_item_douban}
}
\subfigure[Tmall]{
\includegraphics[width=.47\columnwidth]{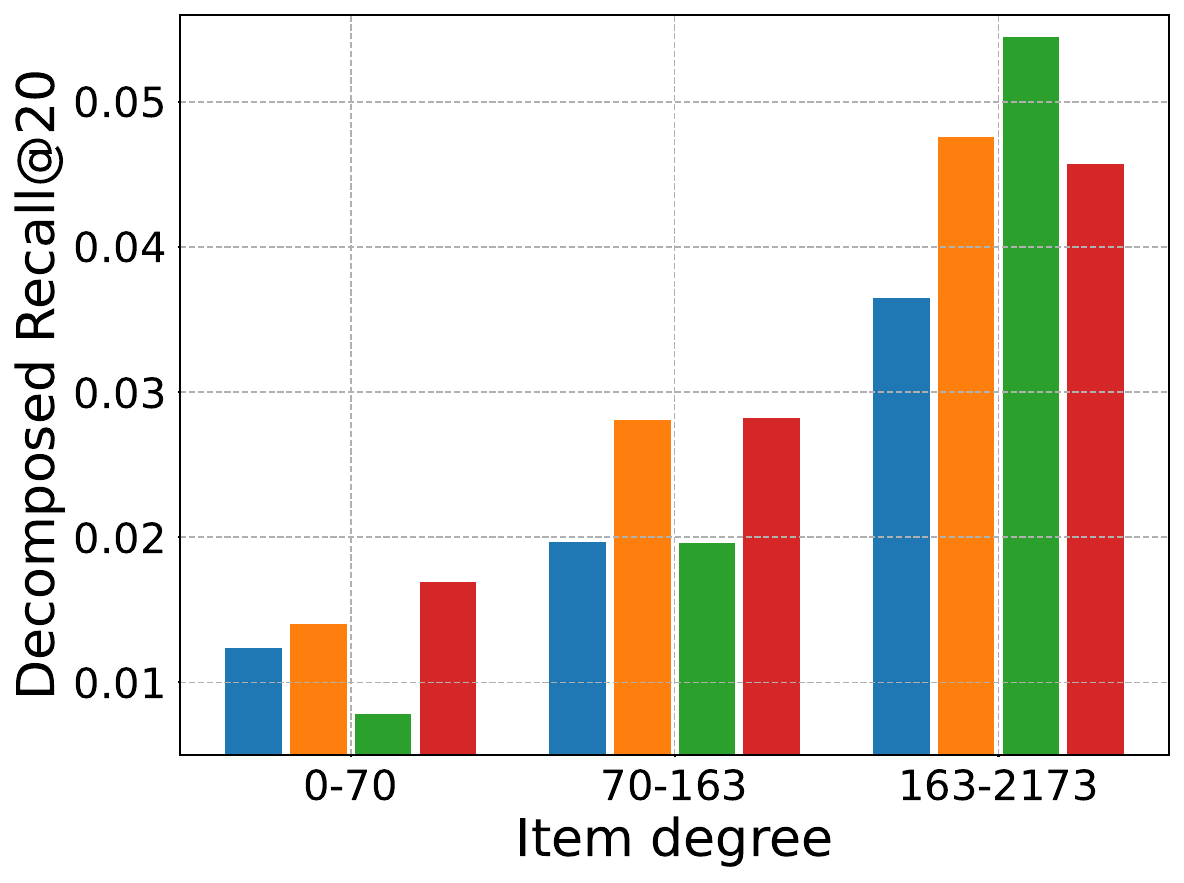}
\label{fig:exp_item_tmall}
}
\caption{Performance comparison over different item groups. More results are in Appendix~\ref{app:robustness}.}\label{fig:exp_item}
\end{figure}

We measure the uniformity of the representation, which is the logarithm of the average pairwise Gaussian potential~\cite{wang2020understanding}:
\begin{align}\label{eq:uniform}
    \mathcal{L}_{\text{uniform}} = \log \underset{\substack{i.i.d \\ u,v \sim p_{\text{node}}}}{\mathbb{E}} e^{-2\|g(u)-g(v)\|_2^2}.
\end{align}where $g(\cdot)$ is the normalized embedding. In Douban and Tmall datasets, we randomly sample 5,000 users and the popular items with more than 200 interactions. After that, we compute the uniformity of their representations in SCONE and baselines with Eq.~\eqref{eq:uniform}. As shown in Fig.~\ref{fig:uniformity}, all uniformity curves show similar trends. In the early steps of training, it has a high uniformity value and then decreases, and after reaching the peak, it tends to gradually converge. In Table~\ref{table:uniformity}, our method shows the highest uniformity at the best performance compared with the other baselines. 
When considered in conjunction with the results in Fig.~\ref{fig:exp_item}, our approach mitigates popularity bias and enhances diversity and personalization.

\begin{table}[t]
\begin{minipage}{0.55\linewidth}
\centering
\includegraphics[width=\columnwidth]{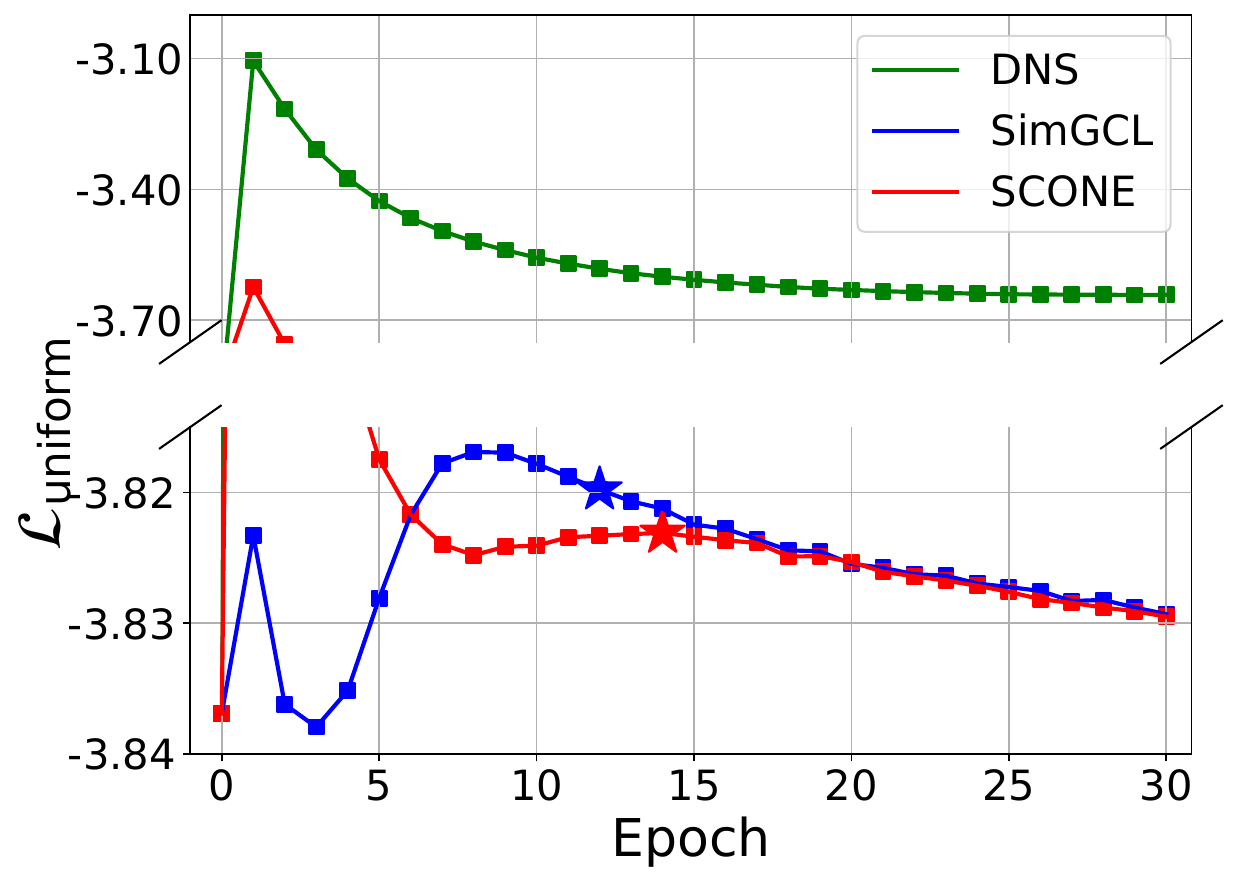}
\vspace{-5pt}
\captionof{figure}{The uniformity curves on Tmall. The stars denote the epoch at the best performance.}
\label{fig:uniformity}
\end{minipage}
\hfill
\begin{minipage}{0.44\linewidth}
\setlength\tabcolsep{1.2pt}
\caption{The uniformity at the best performance. The lower values have higher uniformity.}
\label{table:uniformity}
\begin{center}
\small
\begin{tabular}{lcc}
\toprule
Model & Douban & Tmall \\
\midrule
LightGCN & -2.317 & -3.142\\
SGL & -3.474 & -3.723 \\
SimGCL & -3.469 & -3.820 \\
DNS & -2.639 & -3.642\\
MixGCF & -3.107  & -3.536\\
\midrule
SCONE & \textbf{-3.591}  & \textbf{-3.823}  \\
\bottomrule
\end{tabular}
\end{center}
\end{minipage}
\end{table}

\subsection{Ablation Study of SCONE (RQ3)}\label{subsec:rq3}
\begin{figure}[t]
\centering
\subfigure[Douban]{
\includegraphics[width=.475\columnwidth]{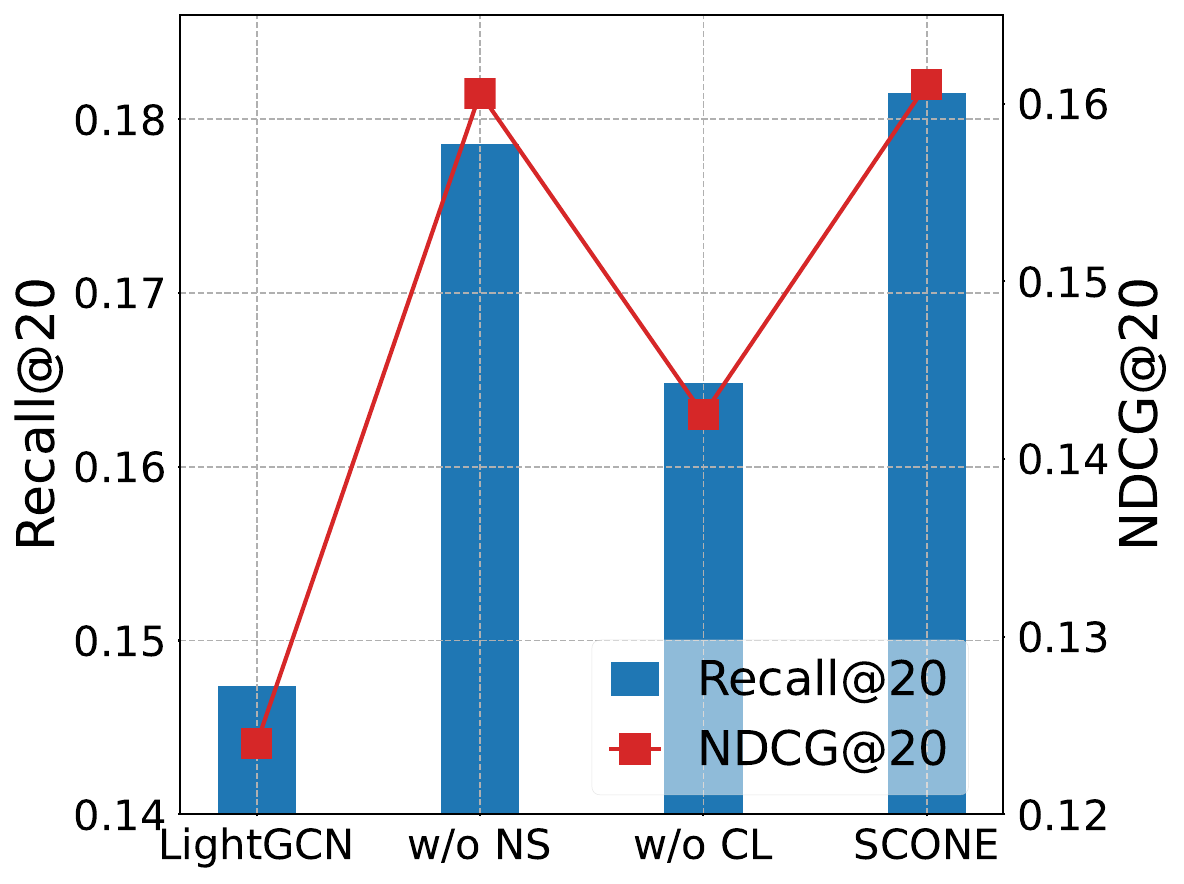}
\label{fig:exp_abl_douban}
}
\subfigure[Gowalla]{
\includegraphics[width=.475\columnwidth]{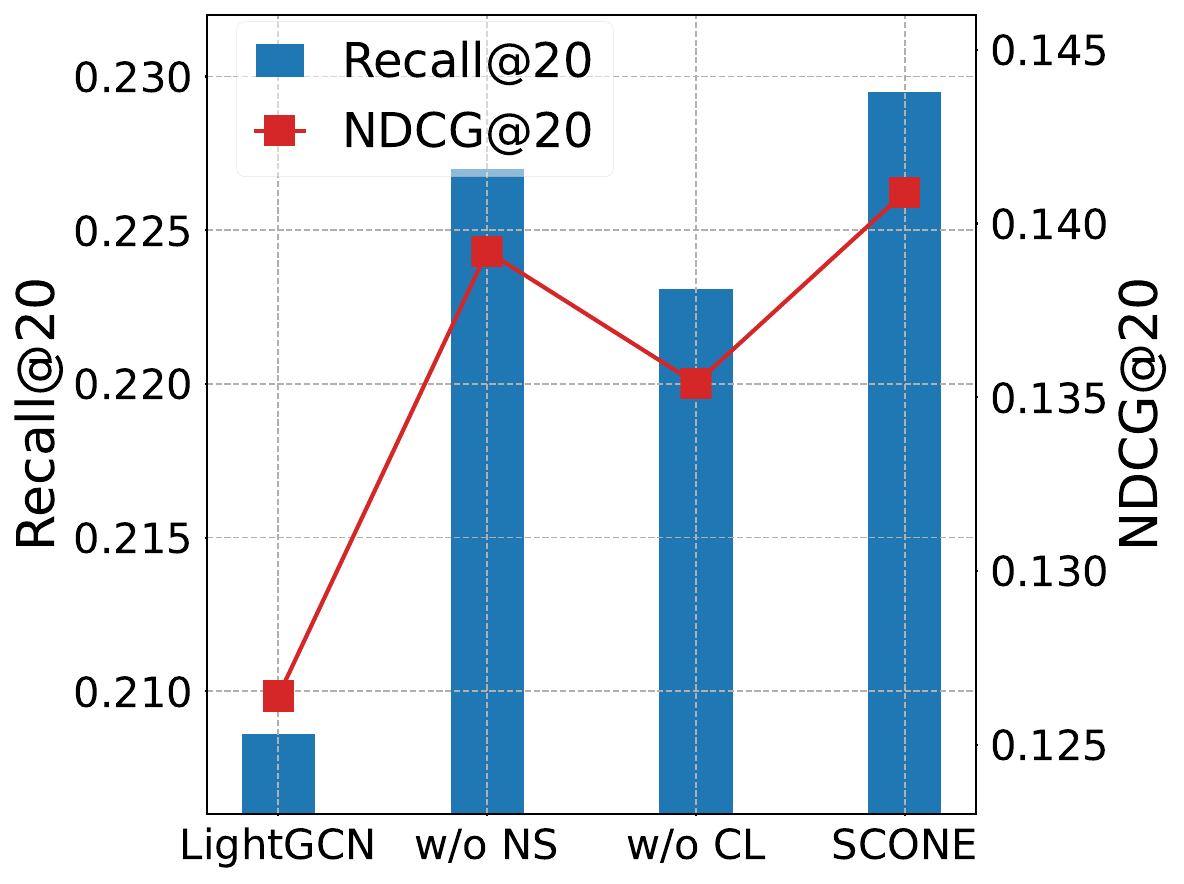}
\label{fig:exp_abl_gowalla}
}
\caption{Ablation study on the efficacy of CL and NS}\label{fig:exp_ablation}
\end{figure}

In Fig.~\ref{fig:exp_ablation}, we conduct ablation studies to evaluate the efficacy of CL and negative sampling in SCONE. `LightGCN' means the model trained without CL and hard negative sampling, `w/o NS' means without hard negative sampling but with only CL, and `w/o CL' means without CL with only hard negative sampling. In all datasets, our CL and hard negative sampling methods improve recommendation performance. 
More results in other datasets are in Appendix~\ref{app:ablation}.

\begin{figure}[t]
\centering
\subfigure[Amazon-CDs]{
\includegraphics[width=.475\columnwidth]{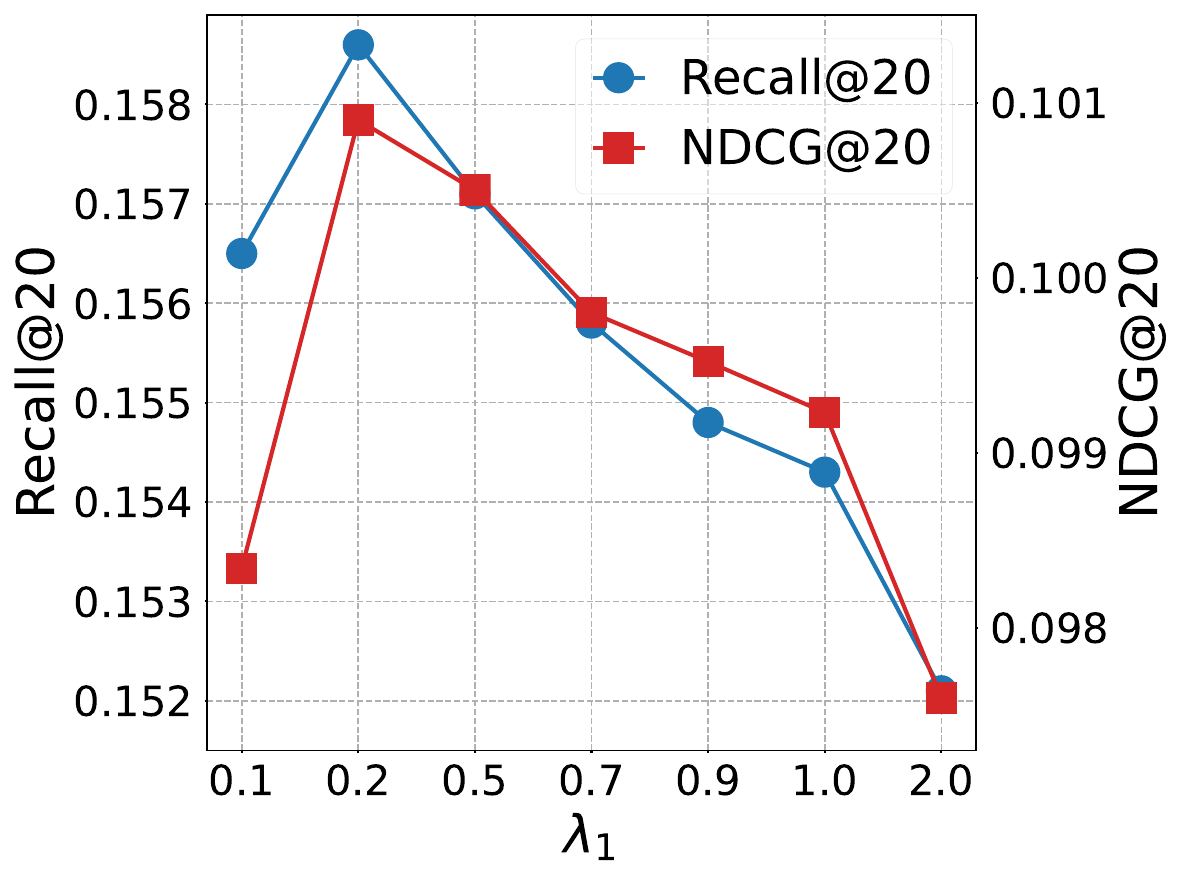}
\label{fig:exp_lambda_amazoncds}
}
\subfigure[ML-1M]{
\includegraphics[width=.475\columnwidth]{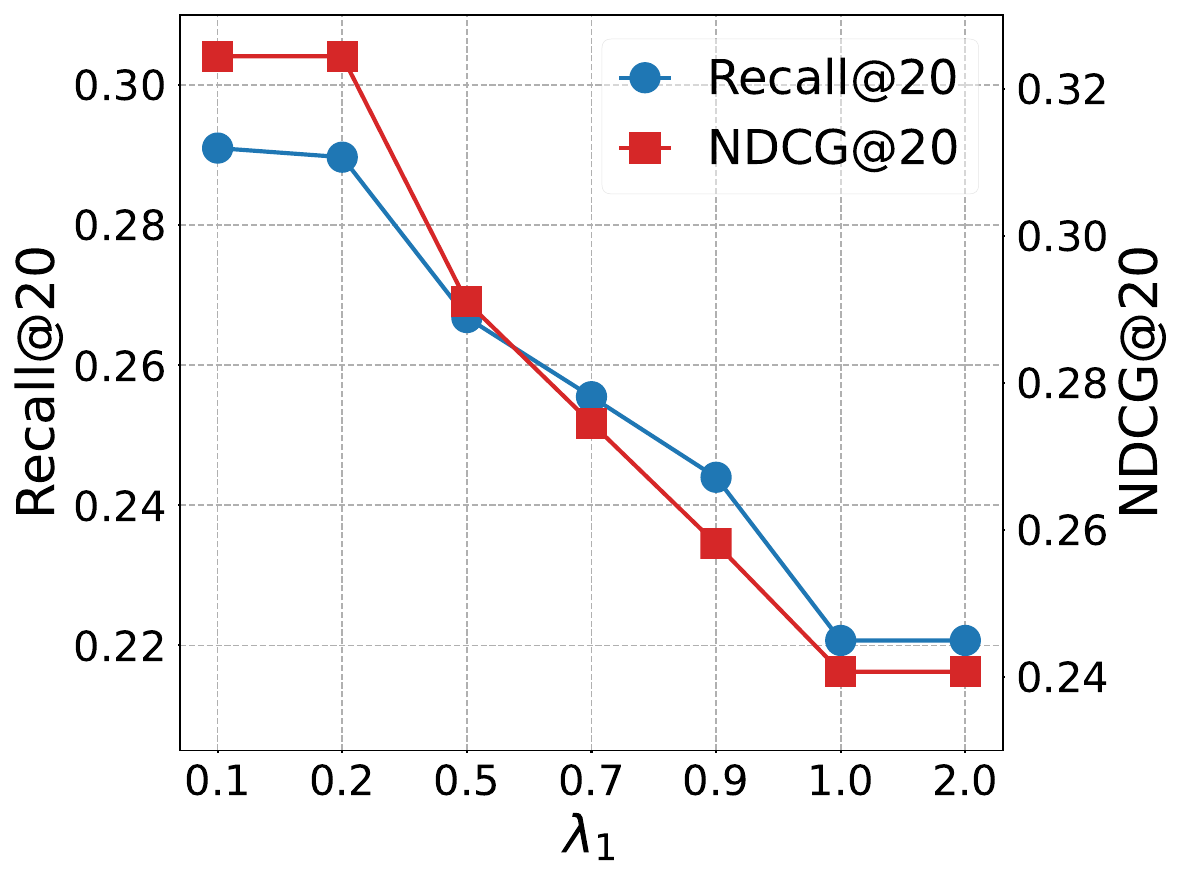}
\label{fig:exp_lambda_ml-1m}
}
\caption{Sensitivity on the regularization weight $\lambda_1$ of CL}\label{fig:exp_lambda}
\end{figure}

\begin{figure}[t]
\centering
\subfigure[Yelp2018]{
\includegraphics[width=.475\columnwidth]{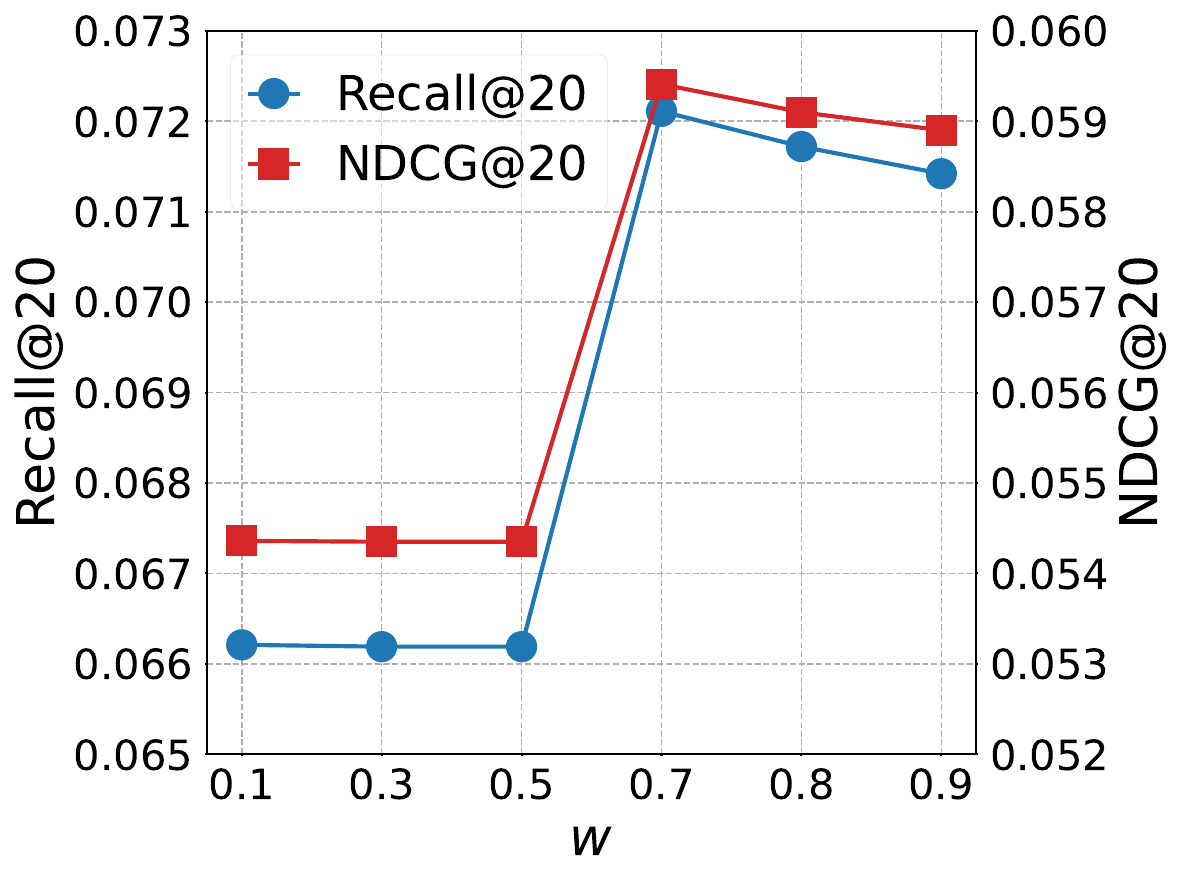}
\label{fig:exp_w_yelp}
}
\subfigure[Amazon-CDs]{
\includegraphics[width=.475\columnwidth]{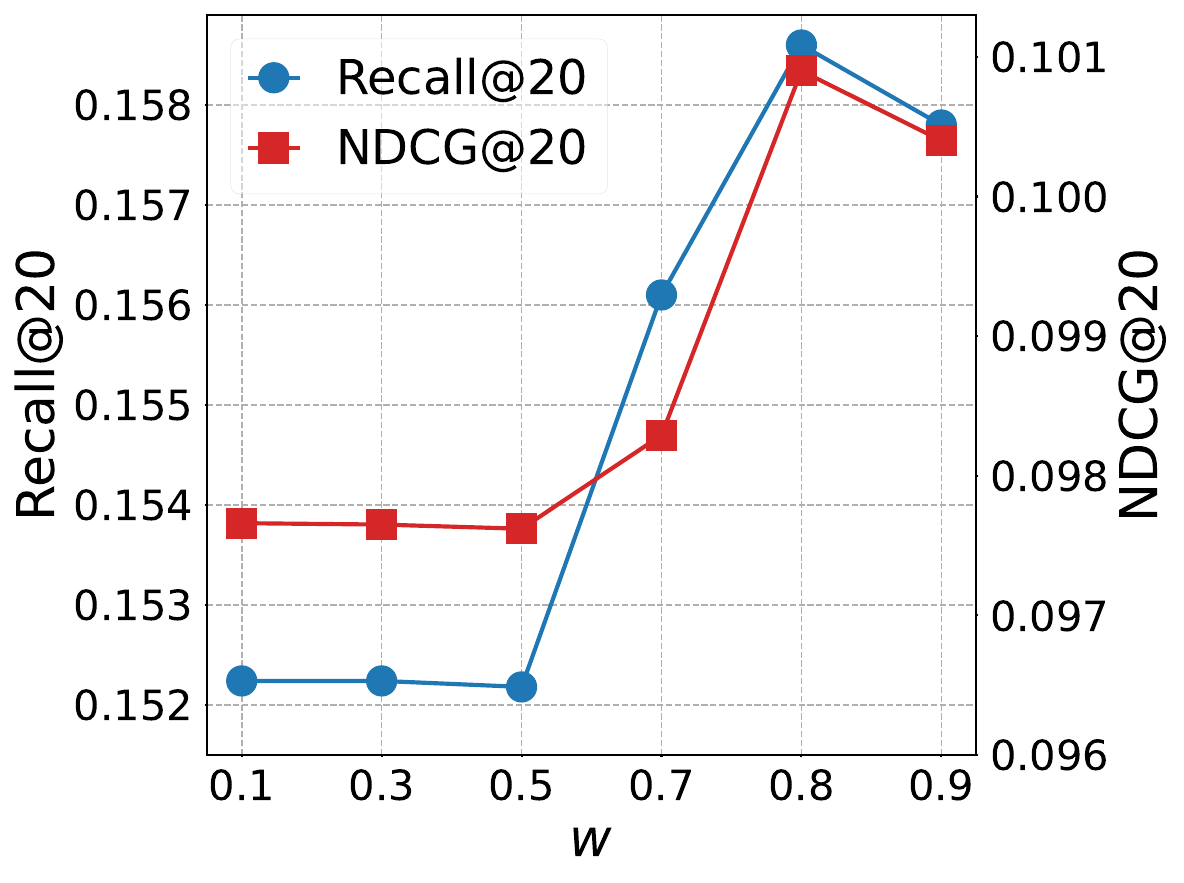}
\label{fig:exp_w_amazoncds}
}
\caption{Sensitivity on the weight $w$}
\label{fig:exp_w}
\end{figure}

\begin{figure}[t]
\centering
\subfigure[Douban]{
\includegraphics[width=.475\columnwidth]{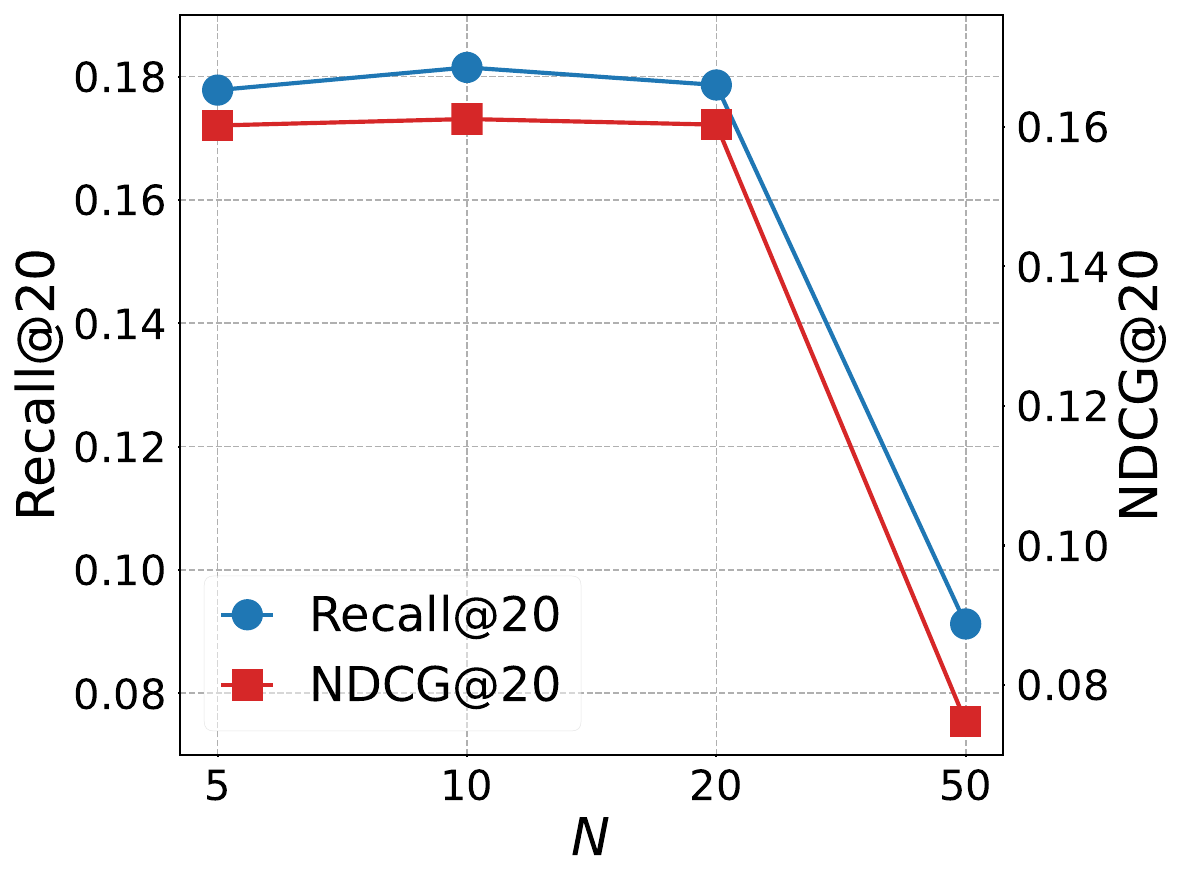}
\label{fig:exp_r_douban}
}
\subfigure[Amazon-CDs]{
\includegraphics[width=.475\columnwidth]{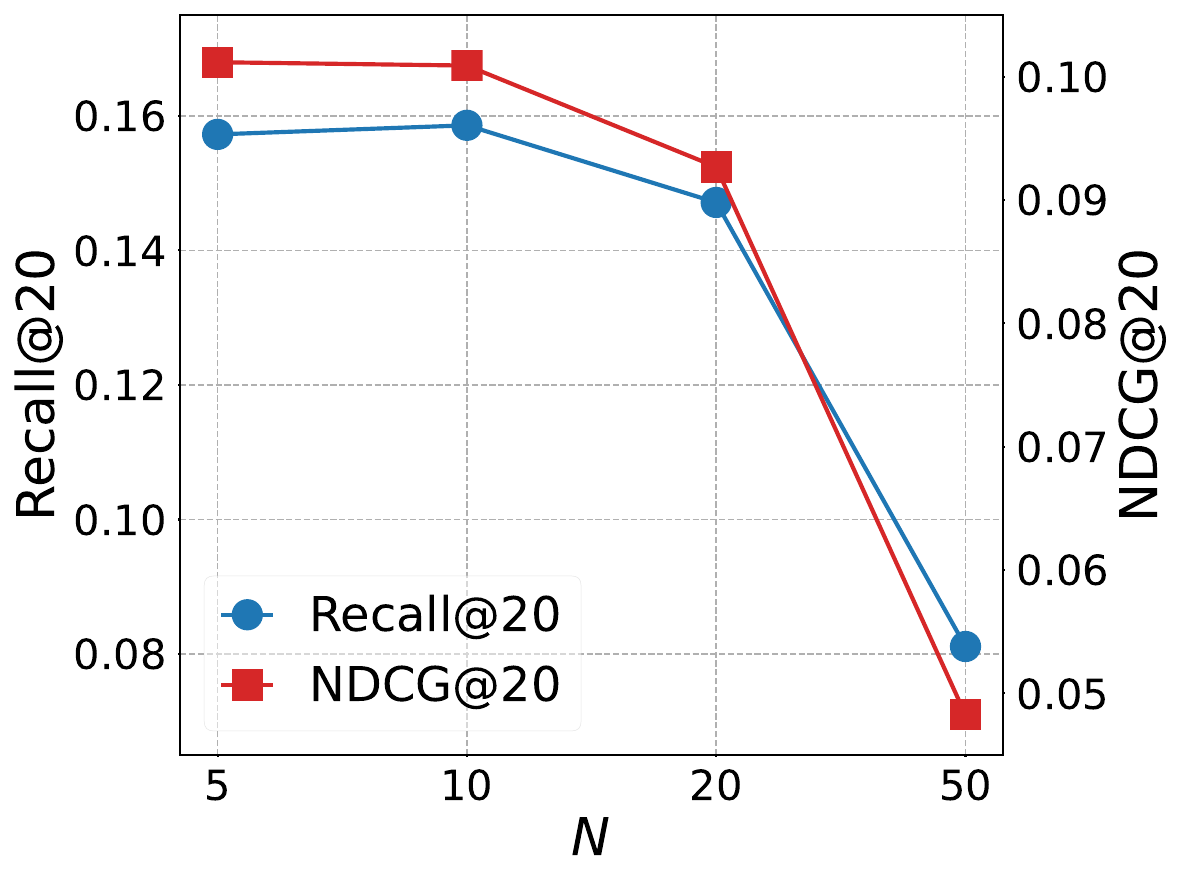}
\label{fig:exp_r_amazoncds}
}
\caption{Sensitivity on the number of sampling steps $N$}
\label{fig:exp_N}
\end{figure}

\subsection{Sensitivity Studies (RQ4)}\label{subsec:rq4}
We analyze the impact of the 3 hyperparameters: the regularization weight $\lambda_1$, the weight $w$ of stochastic positive injection and the number of sampling steps $N$.
\subsubsection{Sensitivity to $\lambda_1$}
Fig.~\ref{fig:exp_lambda} shows the performance of SCONE with varying $\lambda_1$. 
Optimal performance is achieved at $\lambda_1=0.2$ for Amazon-CDs and $\lambda_1=0.1$ for ML-1M, with performance decreasing as $\lambda_1$ increases beyond these points.

\subsubsection{Sensitivity to $w$}
In Fig.~\ref{fig:exp_w}, when $w$ is below 0.7, SCONE shows a dramatic change in performance. This suggests that injecting a moderate amount of positive information while maintaining the characteristics of negative samples is beneficial.

\subsubsection{Sensitivity to $N$}
Fig.~\ref{fig:exp_N} shows model degradation when $N$ exceeds 10 due to excessive perturbation of the embeddings, which weakens user and item-specific characteristics. Thus, we set $N=10$ in all our experiments to balance between augmentation and the preservation of user and item-specific characteristics.

\begin{wraptable}{r}{0.5\columnwidth}
    \vspace{-1.5em}
    \small
    \setlength{\tabcolsep}{1.2pt}
    \centering
    \caption{Runtime (s/epoch)}
    \label{table:runtime}
    \begin{tabular}{lccc}
        \toprule
        Method & Douban & Tmall & Yelp2018  \\
        \midrule
        LightGCN & 3.62s & 47.81s & 15.34s  \\
        SGL & 8.67s  &  122.30s & 37.41s  \\
        SimGCL & 8.72s & 132.82s & 40.04s  \\
        DNS & 25.23s & 154.18s & 71.57s  \\
        MixGCF & 23.25s & 163.78s & 71.29s \\
        SCONE & 19.86s & 142.65s & 60.63s  \\
        \bottomrule
    \end{tabular}
    \vspace{-1em}
\end{wraptable}
\subsection{Runtime Analyses}\label{sec:runtime}
In Table~\ref{table:runtime}, we analyze the runtime per epoch on 3 datasets based on the number of interactions: Douban (small), Yelp2018 (medium), and Tmall (large).
DNS and MixGCF need to compute the negative sampling process. Due to the computational cost of stochastic sampling in diffusion models, 
SCONE takes slightly longer than graph-based CL methods, but SCONE outperforms. 
Compared to existing negative sampling methods, SCONE is faster and shows better performance. 
All baselines and SCONE have the same inference time, which is because the computation time of the final embedding using the graph encoder.

\section{Conclusion and Limitation}
We propose SCONE, a novel framework for addressing data sparsity and negative sampling challenges in graph-based CF models. By using stochastic sampling to generate user-item-specific views and hard negative samples, SCONE performs better on 6 benchmark datasets than on 6 baselines. Our results highlight the robustness of SCONE in handling user sparsity and item popularity issues.

\textbf{Limitations and future work}: 
1) Our scope of focus is LightGCN, leaving adoption of more advanced graph-based CF models for future work. 
2) The task on SCONE differs from sequential recommendation models~\cite{yang2023generate,wu2023diff4rec,liu2023diffusion,niu2024diffusion} using generative models. This distinction allows future work to use the strengths of SCONE in other recommendation tasks.
3) We did not consider more sophisticated techniques such as positive injection of DINS~\cite{wu2023dimension}. Since, our focus was to demonstrate a method capable of simultaneously generating contrastive views and negative samples.
Future work could integrate advanced sampling techniques, extend to various tasks, and apply state-of-the-art GCN and generative models.

\begin{acks}
This work was partly supported by an ETRI grant funded by the Korean government (No. 24ZB1100, Core Technology Research for Self-Improving Integrated Artificial Intelligence System, 70\%), Institute for Information \& Communications Technology Planning \& Evaluation (IITP) grants funded by the Korea government (MSIT) (No. RS-2020-II201361, Artificial Intelligence Graduate School Program (Yonsei University), 10\%; No. RS-2024-00457882, AI Research Hub Project, 10\%), the Korea Advanced Institute of Science and Technology (KAIST) grant funded by the Korea government (MSIT) (No. G04240001, Physics-inspired Deep Learning, 5\%), and Samsung Electronics Co., Ltd. (No. G01240136, KAIST Semiconductor Research Fund (2nd), 5\%)
\end{acks}

\clearpage

\section*{Ethical Considerations}
Collaborative filtering is a technique used to recommend products, services, and contents to users. However, in researching recommendation systems, privacy issues and information bias can arise. As we collect and analyze users' personal data for recommendation, it is important to protect their sensitive information appropriately. In addition, since the recommendation is based on the user's previous behavior, there is a potential for information bias, leading to emphasizing only specific content. Therefore, it should be fairly exposed to various content.


\bibliographystyle{ACM-Reference-Format}
\bibliography{ref}

\clearpage
\appendix

{\huge \textbf{\textit{Appendix}}}

\section{Additional Details for Experiments}\label{app:exp-detail}

\subsection{Datasets}\label{app:dataset}
We provide 6 benchmark datasets used for our experiments. Table~\ref{table:datasets} summarizes the statistical information of the datasets.

\begin{table}[h]
\caption{Statistics of the datasets}
\label{table:datasets}
\begin{center}
\begin{tabular}{lcccc}
\toprule
Dataset & \#Users & \#Items & \#Interactions & Density \\
\midrule
Douban & 12,638 & 22,222 &598,420 & 0.213\% \\
Gowalla &50,821 & 57,440& 1,302,695&0.045\%  \\
Tmall & 47,939 & 41,390&2,619,389 & 0.132\%\\
Yelp2018 & 31,668 & 38,048 & 1,561,406 & 0.130\% \\
Amazon-CDs & 43,169 & 35,648 & 777,426 & 0.051\% \\
ML-1M & 6,038 & 3,492 &  575,281 & 2.728\%  \\
\bottomrule
\end{tabular}
\end{center}
\vspace{-1em}
\end{table}

\subsection{Hyperparameters}\label{app:hyperparameters}
To make a fair comparison with previous studies, we follow hyperparameter settings in the original papers. For all models, the embedding size $D$ is set to 64, the batch size is 2048, and we use $L=2$ for the graph encoder. For our model, we use only VE SDE, and the learning rate is 0.001. We set $\sigma_{\text{min}}=0.01$, $\sigma_{\text{max}}=50$, $T=100$, $N=10$, $\tau=0.2$, $\lambda_2 = 10^{-4}$, $\text{dim}(h_0) = 64$, and $\text{dim}(h_1) = 128$. We tune $\lambda_1$ within the ranges of $\{0.1, 0.2, 0.5, 0.7, 0.9, 1.0, 2.0, 2.5\}$, and $w$ is searched from $\{0.7, 0.8, 0.9\}$. The best hyperparameters in each dataset are as follows: In douban, $\lambda_1 = 0.5 $ and $w=0.9$. In Gowalla, $\lambda_1 = 0.9 $ and $w=0.7$. In Tmall, $\lambda_1 = 2.5$ and $w=0.9$. In Yelp2018, $\lambda_1 = 0.7 $ and $w=0.7$. In Amazon-CDs, $\lambda_1 = 0.2$ and $w=0.8$. In ML-1M, $\lambda_1 = 0.1 $ and $w=0.8$.

\subsection{Experimental environments}
Our software and hardware environments are as follows: \textsc{Ubuntu} 18.04.6 LTS, \textsc{Python} 3.10.8, \textsc{Pytorch} 1.11.0, \textsc{CUDA} 11.7, and \textsc{NVIDIA} Driver 470.161.03, i9, CPU, and \textsc{NVIDIA RTX 3090}.

\section{Additional Results in Robustness of SCONE}\label{app:robustness}
To thoroughly evaluate the robustness of SCONE, we analyze its performance under different user sparsity levels and item popularity groups. This analysis provides insights into how well SCONE handles common challenges in recommendation systems, such as the cold start problem and long-tail item recommendations.

\subsection{User Sparsity}
To evaluate the ability of SCONE to mitigate user sparsity issues, we categorize users into 3 subsets based on their interaction frequency:
\begin{itemize}
    \item Low activity: The bottom 80\% of users with the fewest interactions
    \item Medium activity: Users in the 80th to 95th percentile of interaction frequency
    \item High activity: The top 5\% most active users
\end{itemize}

\begin{figure}[th!]
    \centering
    \subfigure[Gowalla]{
    \includegraphics[width=.47\columnwidth]{img/exp/gowalla_user_sparsity.pdf}
    }
    \subfigure[Tmall]{
    \includegraphics[width=.47\columnwidth]{img/exp/tmall_user_sparsity.pdf}
    }
    \subfigure[Yelp2018]{
    \includegraphics[width=.47\columnwidth]{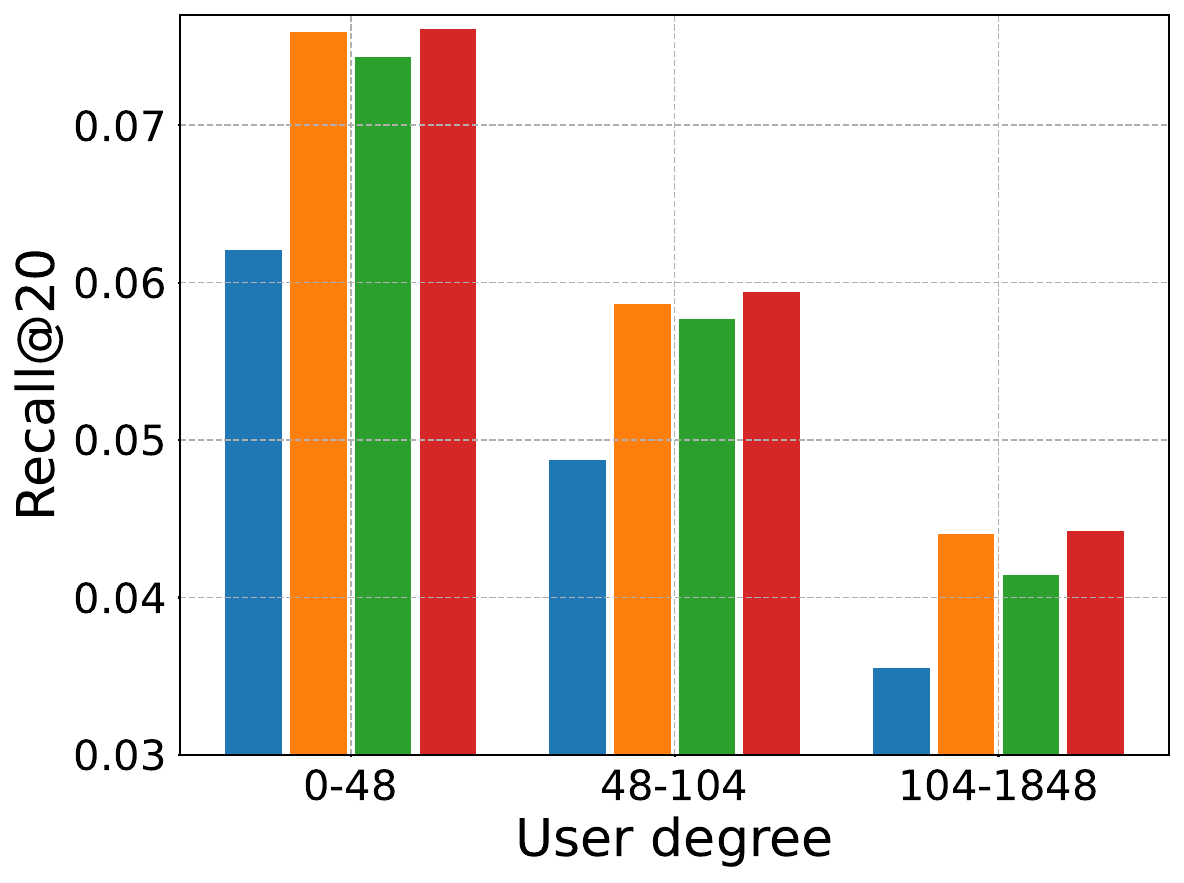}
    }
    \subfigure[Amazon-CDs]{
    \includegraphics[width=.47\columnwidth]{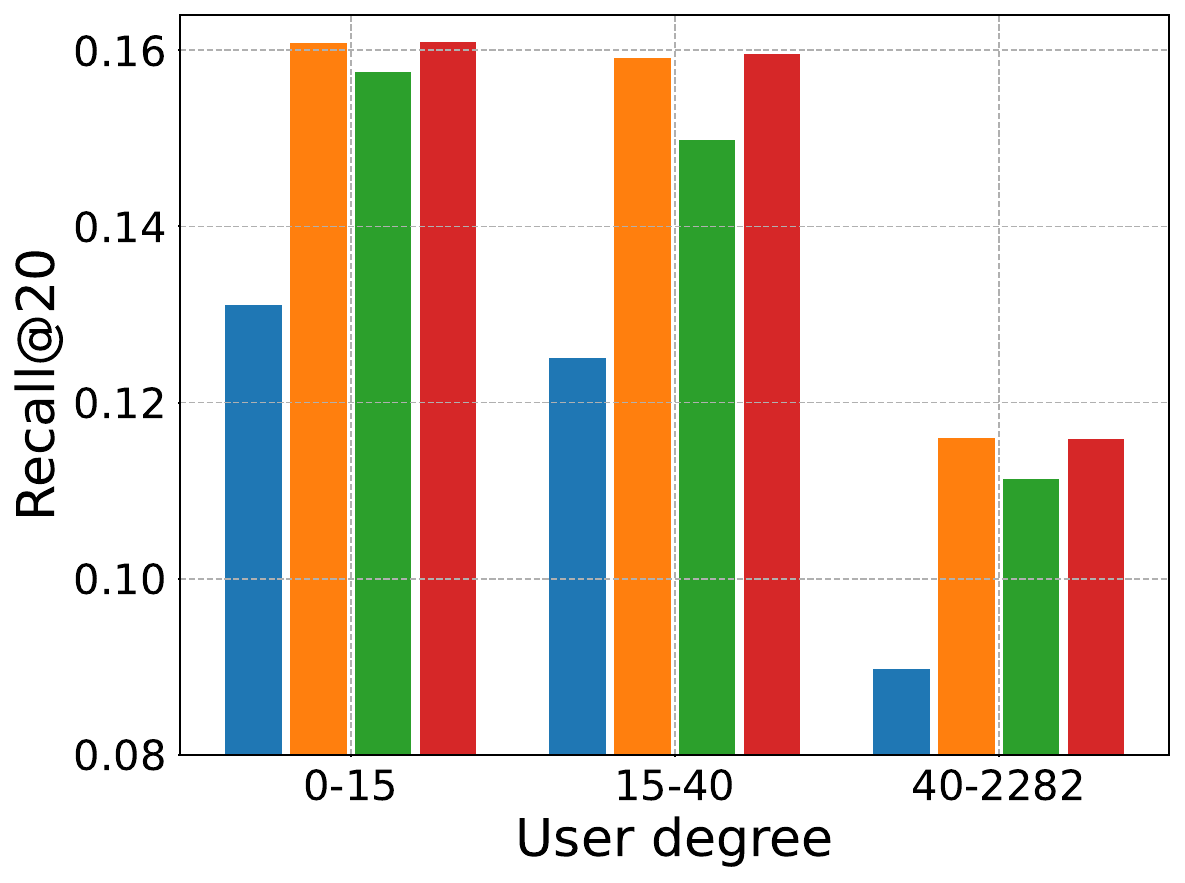}
    }
    \caption{Performance comparison over different user groups}\label{fig:exp_user-app}
\end{figure}

As shown in Fig.~\ref{fig:exp_user-app}, SCONE consistently outperforms other baselines in almost all user groups in the 4 datasets analyzed. In particular, SCONE shows good performance for low activity user groups for all datasets, indicating its effectiveness in addressing the cold start problem for users with limited interaction history.

The performance gap between SCONE and other methods is particularly pronounced in Gowalla and Tmall for low activity users. It shows that our stochastic sampling approach of SCONE is especially beneficial in scenarios with high user sparsity. This robustness can be attributed to the ability of SCONE to generate more informative contrastive views and hard negative samples, even with limited user data.

\subsection{Item Popularity}

To analyze performance of SCONE in different item popularity levels, we use a similar stratification approach:
\begin{itemize}
    \item Low popularity: The bottom 80\% of items with the fewest interactions
    \item Medium popularity: Items in the 80th to 95th percentile of interaction frequency
    \item High popularity: The top 5\% most popular items
\end{itemize}

We use ``decomposed Recall@20'' metrics which represents the contribution of each item group to the overall Recall@20 metric. This allows us to analyze how well SCONE performs for items across the popularity spectrum.

\begin{figure}[th!]
    \centering
    \subfigure[Douban]{
    \includegraphics[width=.47\columnwidth]{img/exp/douban_popularity_bias.pdf}
    }
    \subfigure[Tmall]{
    \includegraphics[width=.47\columnwidth]{img/exp/tmall_popularity_bias.pdf}
    }
    \subfigure[Amazon-CDs]{
    \includegraphics[width=.47\columnwidth]{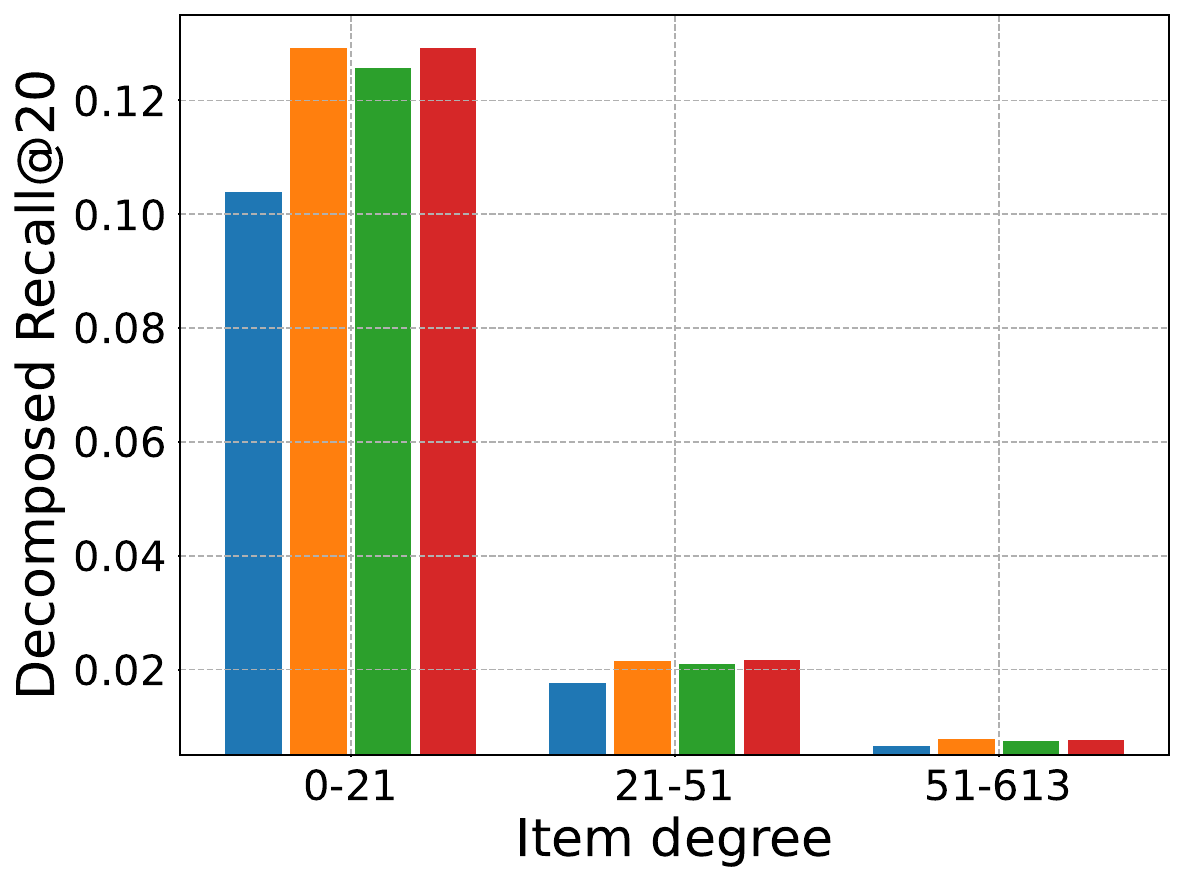}
    }
    \subfigure[ML-1M]{
    \includegraphics[width=.47\columnwidth]{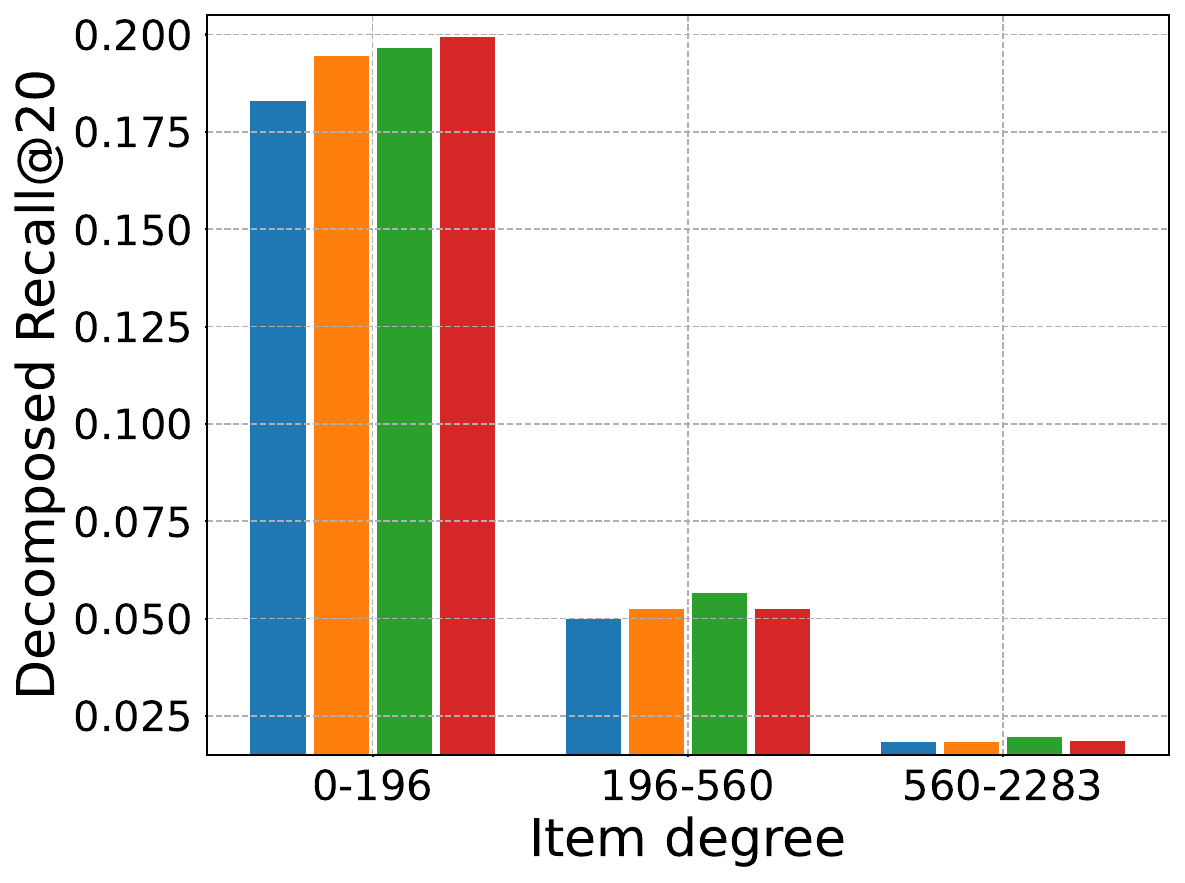}
    }
    \caption{Performance comparison over different item groups}
    \label{fig:exp_item-app}
\end{figure}

As shown in Fig.~\ref{fig:exp_item-app}, SCONE shows notable performance improvements over other baselines, particularly for low popularity items for all 4 datasets. This finding is significant as it indicates the robustness of SCONE in handling the long-tail item recommendation problem.

The performance gain for low popularity items is especially significant in Tmall and Amazon-CDs datasets. This suggests that SCONE is particularly effective in e-commerce scenarios, where the ability to recommend less popular or uncommon items can significantly enhance user experience and potentially increase sales diversity.

\section{Ablation Studies}\label{app:ablation}
The results of our ablation studies, conducted on 4 datasets, show the efficacy of core components of SCONE. 
`LightGCN' means a model trained without CL and hard negative sampling, `w/o NS' means a model trained only with CL and without hard negative sampling, and `w/o CL' means a model trained only with hard negative sampling and without CL. 

\begin{figure}[b!]
    \centering
    \subfigure[Douban]{
    \includegraphics[width=.47\columnwidth]{img/exp/ablation/douban_ablation.pdf}
    \label{fig:exp_abl_douban}
    }
    \subfigure[Gowalla]{
    \includegraphics[width=.47\columnwidth]{img/exp/ablation/gowalla_ablation.pdf}
    \label{fig:exp_abl_gowalla}
    }
    \subfigure[Amazon-CDs]{
    \includegraphics[width=.465\columnwidth]{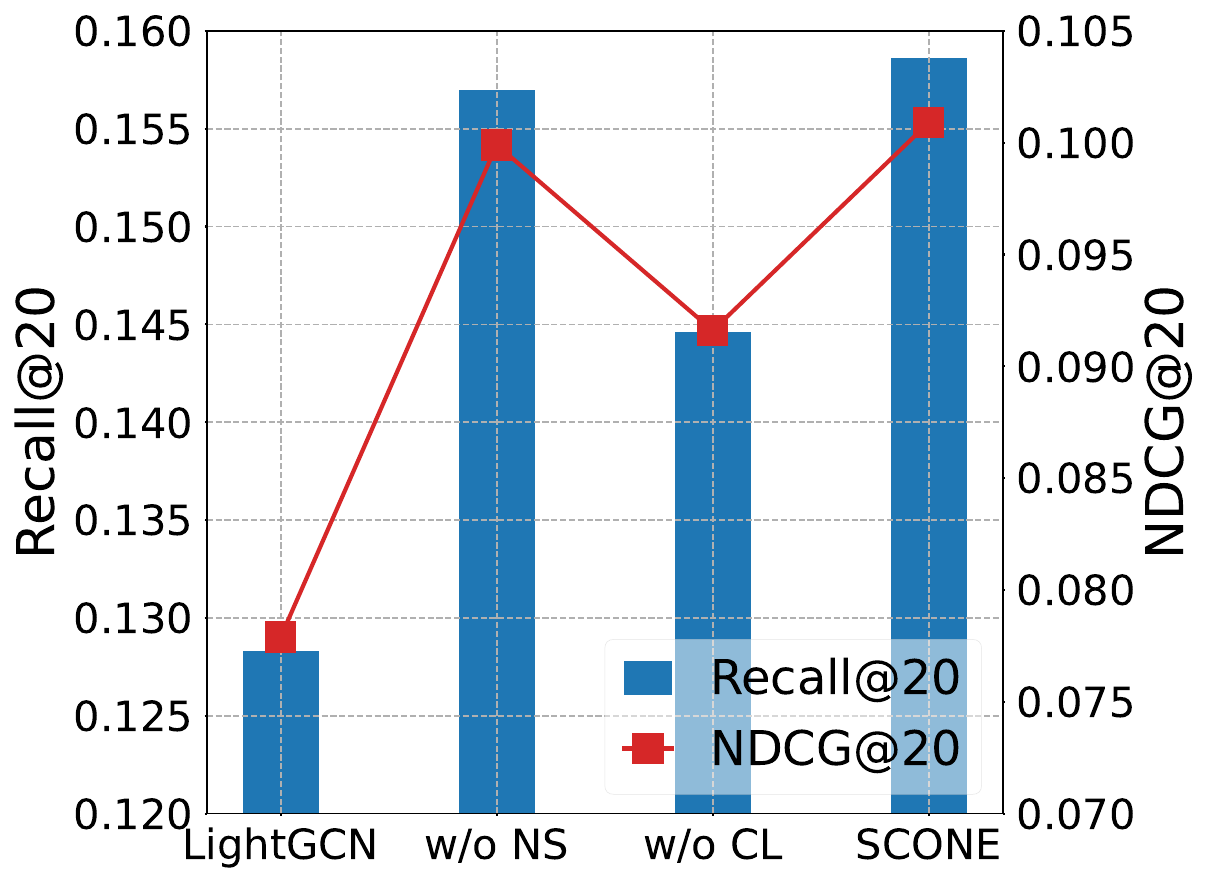}
    \label{fig:exp_abl_amazon}
    }
    \subfigure[ML-1M]{
    \includegraphics[width=.465\columnwidth]{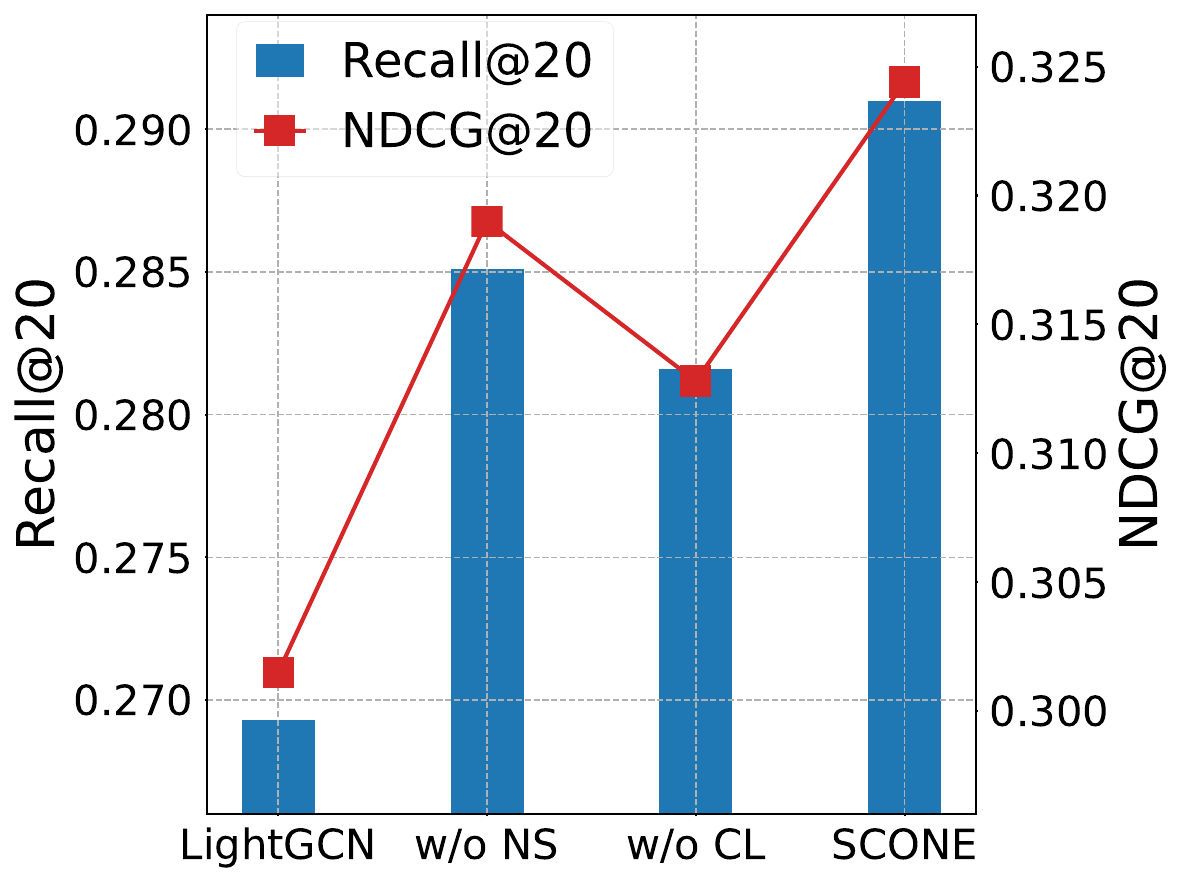}
    \label{fig:exp_abl_ml-1m}
    }
    \caption{Ablation study on the efficacy of the contrastive learning and negative sampling}
    \label{fig:exp_ablation-app}
\end{figure}

Fig.~\ref{fig:exp_ablation-app} compares the performance of `LightGCN', `w/p NS', `w/o CL`, and the full SCONE model.
The results show that both CL and hard negative sampling have a significant positive impact on performance. The full SCONE model outperforms compared to its variants
It is notable that the impact of each component varies on datasets. For instance, CL has a more effect in Douban, while hard negative sampling is more significant in Amazon-CDs. 
Additionally, improvements in NDCG@20 are often more substantial than Recall@20.

\clearpage

\end{document}